\documentclass[twocolumn,trackchanges]{aastex61}
\pdfoutput=1 
\usepackage{amsmath,amstext}
\usepackage[T1]{fontenc}
\usepackage{apjfonts} 
\usepackage[figure,figure*]{hypcap}

\shorttitle{Pair-Instability Supernova Simulations}
\shortauthors{Gilmer et al.} 

\begin{document}

\title{Pair-Instability Supernova Simulations: Progenitor Evolution, Explosion, and Light Curves}

\author{Matthew S. Gilmer}
\affiliation{Department of Physics, North Carolina State University, Raleigh NC, 27695, USA}
\author{Alexandra Kozyreva}
\affiliation{The Raymond and Beverly Sackler School of Physics and Astronomy, Tel Aviv University, Tel Aviv 69978, Israel}
\author{Raphael Hirschi}
\affiliation{Astrophysics group, School of Chemical and Physical Sciences, Keele University, Keele, Staffordshire, ST5 5BG, UK}
\affiliation{Kavli Institute for the Physics and Mathematics of the Universe (WPI), The University of Tokyo Institutes for Advanced Study, The University of Tokyo, 5-1-5 Kashiwanoha, Kashiwa, Chiba 277-8583, Japan}
\author{Carla Fr\"{o}hlich}\affiliation{Department of Physics, North Carolina State University, Raleigh NC, 27695, USA}
\author{Norhasliza Yusof}
\affiliation{Department of Physics, Faculty of Science, University of Malaya, 50603 Kuala Lumpur, Malaysia}
\email{msgilmer@ncsu.edu}

\begin{abstract}
In recent years, the viability of the pair-instability supernova (PISN) scenario for explaining superluminous supernovae has all but disappeared except for a few slowly-evolving examples. However, PISN are not predicted to be superluminous throughout the bulk of their mass range. In fact, it is more likely that the first PISN we see (if we have not seen one already) will not be superluminous. Here, we present hydrodynamic simulations of PISNe for four stellar models with unique envelope properties spanning the PISN mass range. In addition, we compute synthetic light curves for comparison with current and future observations. We also investigate, in the context of our most massive model, the prospect of mixing in the supernova ejecta alleviating discrepancies between current PISN models and the remaining superluminous candidate events. To this end, we present the first published 3D hydrodynamic simulations of PISNe. After achieving convergence between 1D, 2D, and 3D simulations we examine mixing in the supernova ejecta and its affect on the bolometric light curve. We observe slight deviations from spherical symmetry which increase with the number of dimensions. We find no significant effects on the bolometric light curve, however we conclude that mixing between the silicon and oxygen rich layers caused by the Rayleigh-Taylor instability may affect spectra.
\end{abstract}


\section{Introduction}

Pair-instability supernovae (PISNe) are the explosive deaths of very massive stars (VMS; defined by \cite{Vink2014} as stars with initial masses greater than 100~M$_{\odot}$) that produce carbon-oxygen (CO) cores in the mass range 60~M$_{\odot} \lesssim$ M\textsubscript{CO} $\lesssim$ 130~M$_{\odot}$. Stellar models predict that, for non-rotating stars with zero metallicity, this corresponds to a zero-age main sequence (ZAMS) mass range of 140~M$_{\odot}$ $<$ M\textsubscript{ZAMS} $<$ 260~M$_{\odot}$ \citep{HW02}. For stars in this mass range, life is cut short when the pair-instability (PI) triggers an implosion of the core shortly after core carbon burning. The implosion is reversed by explosive nuclear burning (of primarily oxygen) which releases enough energy to totally unbind the star. The PI occurs when the radiation pressure in the stellar core is reduced by the reaction $\gamma + \gamma \rightarrow e^- + e^+$ and was first shown to cause explosions in simulations of isentropic oxygen cores \citep{Rakavy1967,Barkat1967}. At the time, detection of such an event was thought to be highly unlikely since massive enough progenitors within the range of detectability were thought to be extremely rare, if they existed at all. Since then, observations of VMS along with advances in stellar physics have made the search much more promising. 

A few VMS, have been detected in the Large Magellanic Cloud, specifically in the cluster R136 \citep{Crowther2010}. In addition, there are several good VMS candidates near the Galactic center \citep{Martins2014}. However, it is unclear whether or not these stars will be able to retain enough mass to explode as PISNe. They exist in regions where the metallicity is near solar which is thought to drive very high mass-loss rates. \cite{Langer2007} find a limiting metallicity of Z$_{\odot}$/3 above which no star will be able to explode as a PISN. In any case, their existence suggests that similarly massive stars have formed in regions of lower metallicity and in the early Universe.

Including rotation leads to a more chemically homogeneous stellar evolution allowing more of the initial mass to be converted into heavier elements. This effect facilitates the formation of larger carbon-oxygen cores for a given ZAMS mass, shifting the minimum ZAMS mass down from 140~M$_{\odot}$ to 65~M$_{\odot}$ for stars with initial rotation rates of 80\% of Keplerian velocity \citep{Chatzopoulos2012}. In other words, in a model for a 65~M$_{\odot}$ star rotating at this velocity, all of its initial mass was converted into carbon and oxygen. Additionally, the presence of a magnetic field at the surface of a VMS can quell mass loss rates allowing for the possibility of PISNe even from solar-metallicity progenitors \citep{Georgy2017}. These results substantially lower the bar for finding PISN progenitor candidates.

Many recent numerical simulations agree that the Pop~III Initial Mass Function (IMF) is dominated by stars around 100~M$_{\odot}$ \citep{Abel2002,Bromm2004,Yoshida2008}. However, some very recent simulations find that fragmentation may lead to stellar populations that extend below even 1~M$_{\odot}$ \citep{Stacy2014,Stacy2016} which means VMS would be significantly less abundant than previously thought. Additionally, it is unclear how massive an individual Pop~III star can get. The maximum mass may be limited by ionization feedback \citep{Krumholz2014}. If the maximum mass is as high as it appears to be in the local Universe, then VMS are likely to exist in the early Universe. Furthermore, such stars would experience lower mass-loss rates due to the absence of metals in their atmospheres, allowing them to retain enough mass to explode as PISNe. There is also numerical evidence that suggests pockets of pristine star-forming gas exist even at relatively low redshift ($2<z<5$) \citep{Tornatore2007} which increases the prospects of PISN progenitor stars existing in the local Universe.

PISNe are expected to produce a wide variety of SN types as well as span a large range in peak luminosity. On the lower end of the PISN mass range, the explosion energies can be only a few Bethe (1~Bethe equals $10^{51}$~erg) and the nickel yields may be less than those of ordinary core-collapse supernovae. Consequently, the light curves (LCs) and spectra may resemble those resulting from other supernova mechanisms. For example, red supergiant and stripped core PISN progenitors would likely look like long-duration luminous Type II-P and Type Ib/Ic SNe, respectively \citep{Kasen2011,Kozyreva2014b}. Low-mass PISNe may even explain some ``.Ia supernovae'' (named as such because their explosion strength is 1/10th that of ordinary supernovae) \citep{Whalen2014}. 

Conversely, near the upper end of the PISN mass range the explosions can be extremely energetic. The yield of radioactive nickel in such explosions can approach 55~M$_{\odot}$ \citep{HW02} causing a very luminous long-duration SN. It is this potential for high luminosity that made high-mass PISNe an attractive model for Superluminous Supernovae (SLSNe) when they (SLSNe) were first observed about a decade ago.

SLSNe are defined as any SN with a peak absolute magnitude brighter than -21 \citep{Gal-Yam2012}. They are classified as either SLSN-I (for events without hydrogen) or SLSN-II (for events with hydrogen) just as in the classification scheme for normal SNe. As is the case for normal SNe, SLSNe display a large amount of diversity within the two main types. The PISN model is most well-suited to explain the slowest evolving SLSNe-I that exhibit post-peak decline rates consistent with the radioactive decay of $^{56}$Ni to $^{56}$Co and $^{56}$Fe \citep{2017MNRAS.468.4642I,Jerkstrand2017}.

\cite{Gal-Yam2009} proposed a PISN explanation for such a SLSN, namerly SN~2007bi (although \cite{Woosley2007} had made the case that SN 2006gy was produced by circumstellar interaction with shells originating from the pulsational pair-instability). SN~2007bi was discovered in a relatively nearby dwarf galaxy ($z=0.1279$) which means that, if it was a PISN, VMS can form and retain enough mass to explode as PISNe in the local Universe. The PISN interpretation of SN~2007bi was both supported \cite{Kasen2011,Kozyreva2014b} and critiqued \cite{Dessart2012,Dessart2013,Chatzopoulos2015,Jerkstrand2016} by subsequent works. PISN models were able to sufficiently match key observables such as the bolometric light curve and photospheric velocity, however they were not able to explain the blue nebular spectra of SN~2007bi.

More recently, PTF12dam has captured the interest of PISN enthusiasts. PTF12dam has late-time \citep{Nicholl2013} and host galaxy \citep{Chen2015} properties that are very similar to those of SN~2007bi but, unlike SN~2007bi, it was caught before peak luminosity \citep{Quimby2012}. The relatively fast rise to peak luminosity together with the spectral evolution over this period pose serious problems for PISN models \citep{Nicholl2013}. However, as shown by \cite{Kozyreva2017} with two of the  models used here (P200 and P250), stripped-envelope PISN models at relatively high metallicity ($Z = 0.001$) predict shorter rise times and higher color temperatures than their Pop~III cousins. Here, we extend the mass range to include two lower mass models (P150 and P175) to cover the PISN mass range. In addition, we extend the dimensionality of our simulations to 2D and 3D in order to examine the effects of mixing of $^{56}$Ni in the SN ejecta and its observational consequences.

We describe our methods in Section~\ref{sec:setup} including the stellar evolution, hydrodynamic (in 1D, 2D, and 3D) and radiation-hydrodynamic simulations. In Section~\ref{sec:flash}, we present the explosion properties from the hydrodynamic simulations and in Section~\ref{sec:stella} we discuss the lightcurves. Finally, we summarize the main points of the paper in Section~\ref{sec:summary}.

\section{Inputs and Numerical Setup}
\label{sec:setup}

\subsection{Stellar Models}

Four VMS models were computed with the \verb|GENEC| stellar evolution code \citep{Ekstrom2012} and with the same input physics as in \citet{Yusof2013}. 
GENEC uses adaptive spatial and temporal resolutions. Spatial resolution is set to resolve gradients of key quantities like temperature and hydrogen content. Around 200 zones are used on the ZAMS and 300-800 zones are used towards the end of the evolution (more zones are used for more extended/cooler envelopes). Models took between 25,000 and 30,000 time steps from start to end.
Mass loss strongly affects the evolution of very massive stars. We therefore list the prescriptions used to calculate the models in this study. For main-sequence stars, we used the prescription for radiative line driven winds from \citet{VN01}, which compares rather well with observations \citep{PAC10,muijres11}. 
For stars in a domain not covered by the \citet{VN01} prescription ($\log (T_\text{eff}) < 3.9$), we applied the \citet{deJager88} prescription to models with $\log (T_\text{eff}) > 3.7$. For $\log (T_\text{eff}) \leq 3.7$, we performed a linear fit to the data from \citet{sylvester98} and \citet{vloon99} \citep[see][]{Crowther01}. The formula used is given in Eq.\,2.1 in \citet{BHP12}.
In the calculations, we consider a transition from O-type or giant to Wolf-Rayet (WR) star when the surface hydrogen mass fraction, $X_s<0.3$ and the effective temperature, $\log (T_\text{eff}) > 4$. 
The mass loss rate used during the WR phase depends on the WR sub-type. For the eWNL phase (when $0.3>X_s>0.05$), the 
\citet{GH08} recipe is used (in the validity domain of this prescription, which usually covers most of the eWNL phase). In many cases, the WR mass-loss rate of \citet{GH08} is lower than the rate of \citet{VN01}, in which case, we used the latter. For the eWNE phase -- when $0.05>X_s$ and the ratio of the mass fractions of $({^{12}\text{C}}+{^{16}\text{O}})/{^{4}\text{He}}<0.03$
-- and WC/WO phases -- when 
$({^{12}\text{C}}+{^{16}\text{O}})/{^{4}\text{He}}>0.03$ -- we used the corresponding prescriptions of \citet{NL00}. Note also that both the \citet{NL00} and \citet{GH08} mass-loss rates account for clumping effects \citep{muijres11}.

The metallicity dependence of mass loss rates is included in the following way. The mass loss rate used at a given metallicity, $\dot{M}(Z)$, is the mass loss rate at solar metallicity, $\dot{M}(Z_\odot)$, multiplied by the ratio of the metallicities to the power of $\alpha$: $\dot{M}(Z)= \dot{M}(Z_\odot)(Z/Z_\odot)^\alpha$, where
$\alpha$ was set to 0.85 for the O-type phase and WN phase and 0.66 for the WC and WO phases following \citet{EV06}.
Note that for WR stars the initial metallicity rather than the actual metallicity was used in the equation above.
The parameter $\alpha$ was set to 0.5 for the \citet{deJager88} prescription. Finally, $\alpha$ was set to 0 (no dependence) if $\log (T_\text{eff}) \leq 3.7$ (note that none of the models presented in this study reach such low effective temperatures).

All models are non-rotating and have an initial metallicity of $Z=0.001$. Considering solar composition to be $Z=0.014$, this means that the initial metallicity $Z\simeq 0.07\,Z_\odot$.
Their ZAMS masses are given in the names of the models (P150, P175, P200, and P250) and their pre-SN properties are listed in Table \ref{tab:models}. The masses were chosen to span the PISN mass range given in \cite{HW02}, that is 140 M$_{\odot}$ $<$ M$_{\mathrm{ZAMS}}$\,$<$\,260\,M$_{\odot}$. 

\begin{table}[tbh]
\centering
\caption{Pre-SN properties for models P150, P175, P200, and P250. 
    }
\label{tab:models}
\begin{tabular}{lllllllll}
\tableline
   \multicolumn{1}{l}{Model}
   & \multicolumn{1}{l}{$M_{\text{tot}}$} & \multicolumn{1}{l}{$M_{\text{CO}}$} & \multicolumn{1}{l}{Radius } & \multicolumn{1}{l}{Surface Composition} \\
& \multicolumn{1}{l}{(M$_{\odot}$)} & \multicolumn{1}{l}{(M$_{\odot}$)} & \multicolumn{1}{l}{(R$_{\odot}$)} & \multicolumn{1}{l}{ } \\
\tableline
P150 & 90.8 & 65.7 & 1267 & 20\% H; 80\% He  \\
P175 & 102.8 & 81.4 & 1107 & 18\% H; 82\% He \\
P200 & 109.9 & 100.9 & 80.1 & 6\% H; 94\% He \\
P250 & 126.7 & 126.7 & 2.4 & 34\% He; 39\% C; 27\% O \\
\tableline
\end{tabular}
\tablecomments{
	The CO mass core is defined as the mass coordinate at which the sum of the carbon and oxygen mass fractions falls below 0.5.}
\end{table}

\begin{figure*}
\centering
\begin{tabular}{cc}
\includegraphics[width=0.5\textwidth,clip=]{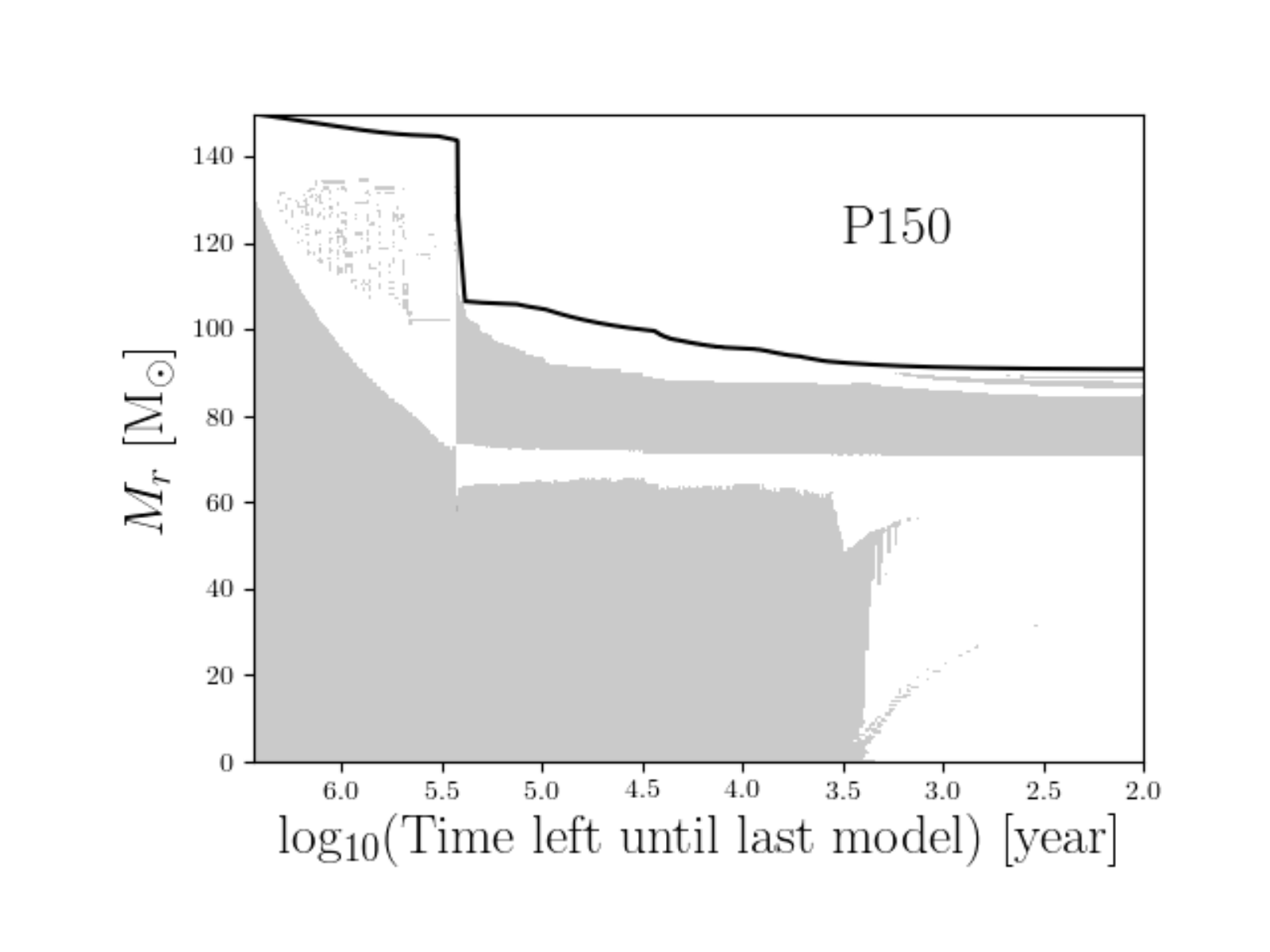} &
\includegraphics[width=0.5\textwidth,clip=]{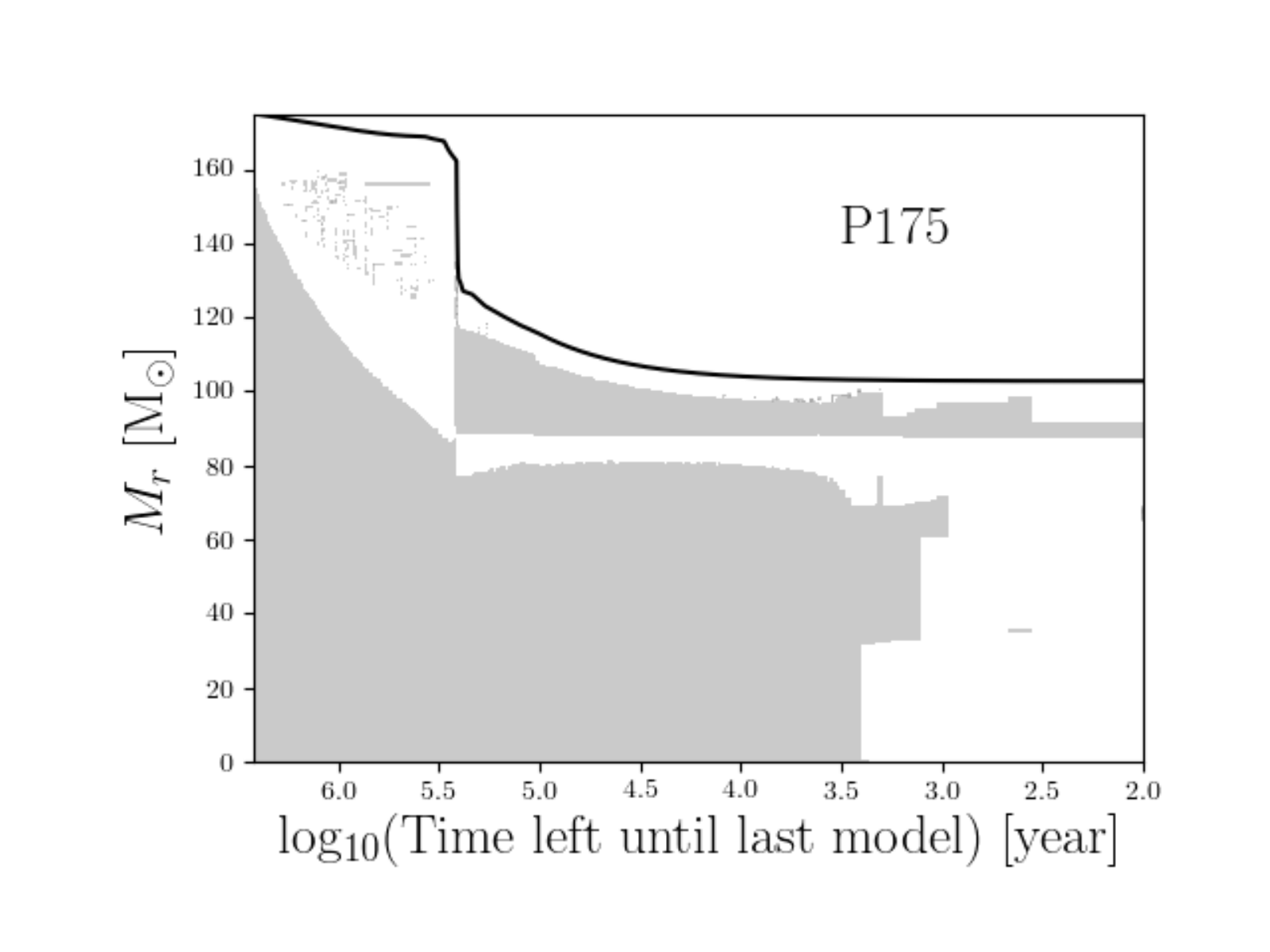} \\
\includegraphics[width=0.5\textwidth,clip=]{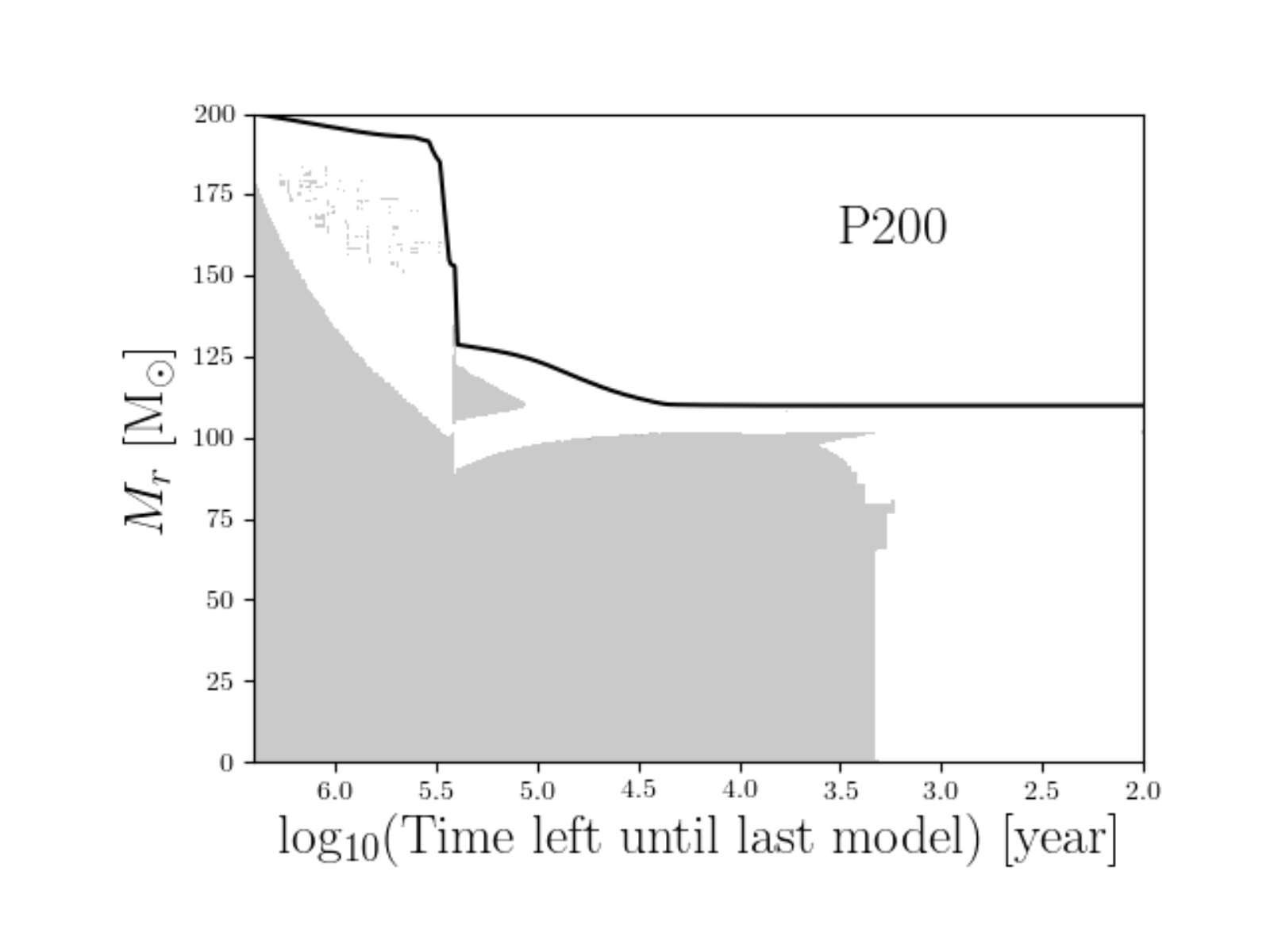} &
\includegraphics[width=0.5\textwidth,clip=]{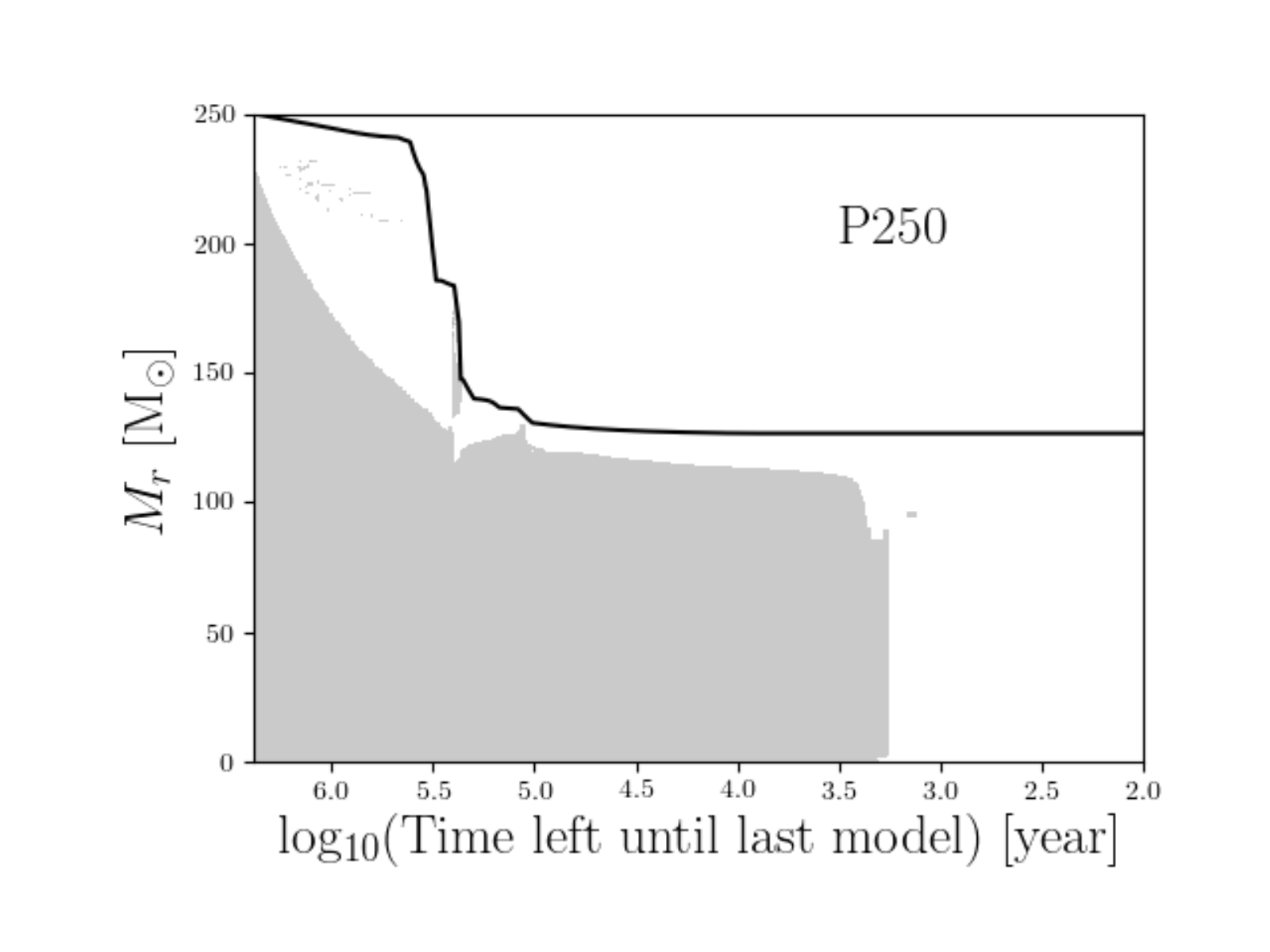} \\
\end{tabular}
\caption{Structure evolution diagram for the VMS models as a function of the log of the time left until the last model. The grey zones 
represent the convective regions. The top solid line corresponds to the total mass. Reddish area indicates the regions where energy is released via nuclear
burning, and bluish area indicates cooling via neutrino losses.}\label{fig:kip_logt}
\end{figure*}

\begin{figure}
\centering
\includegraphics[width=0.5\textwidth]{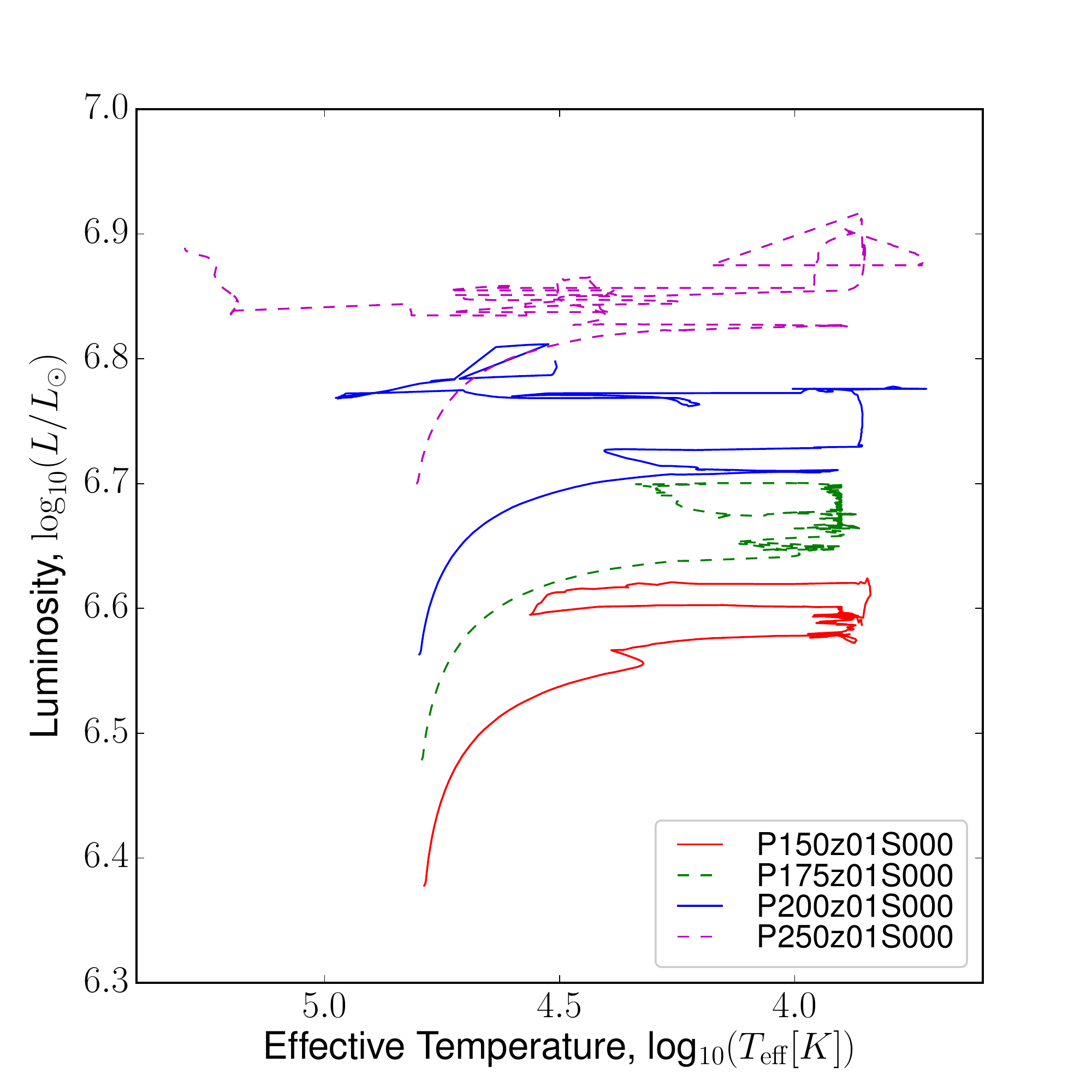}
\caption{Evolutionary tracks in the HR diagram. 
}
\label{fig:hrd}
\end{figure}

\begin{figure*}
\centering
\includegraphics[width=0.5\textwidth]{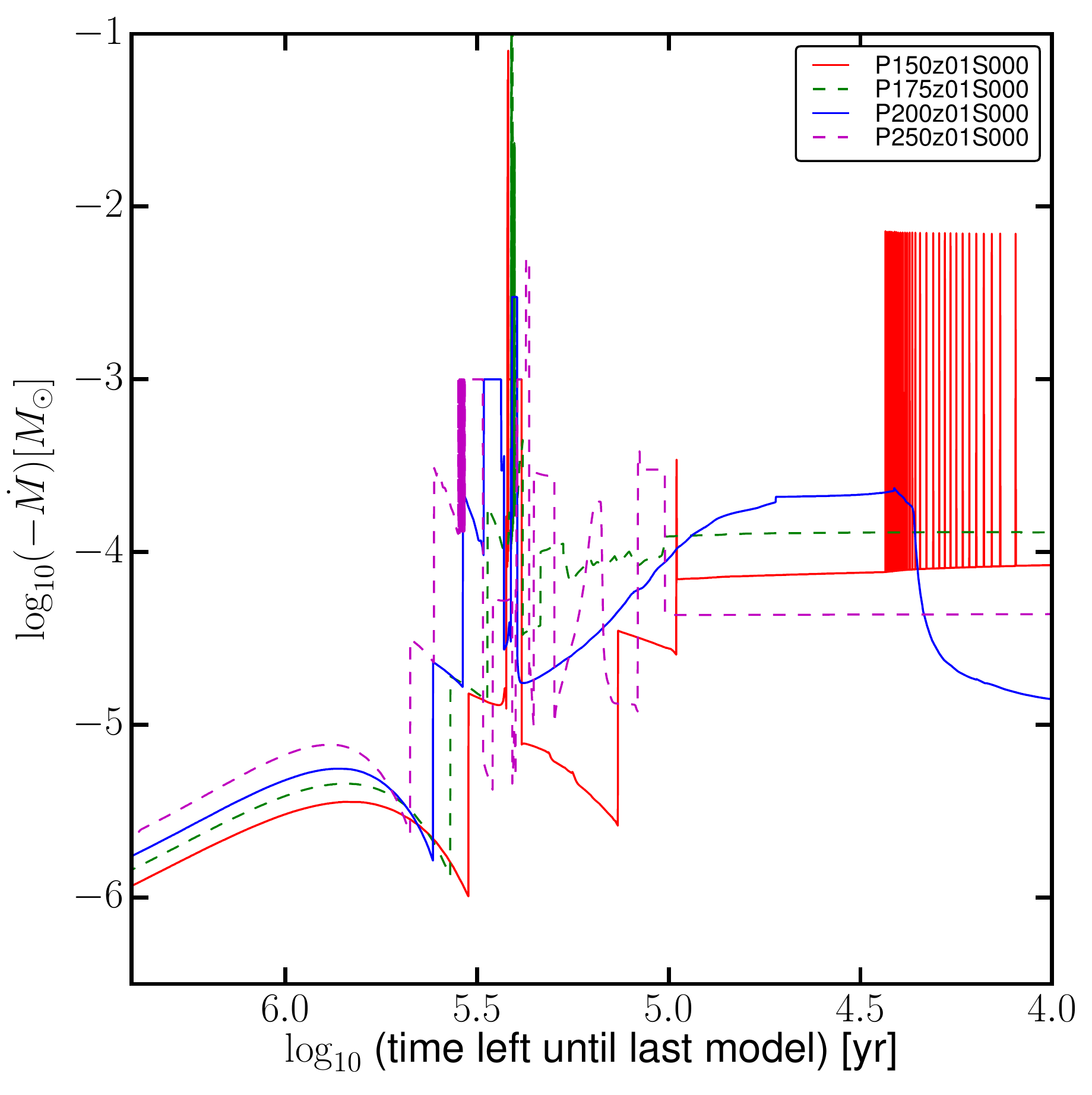}\includegraphics[width=0.5\textwidth]{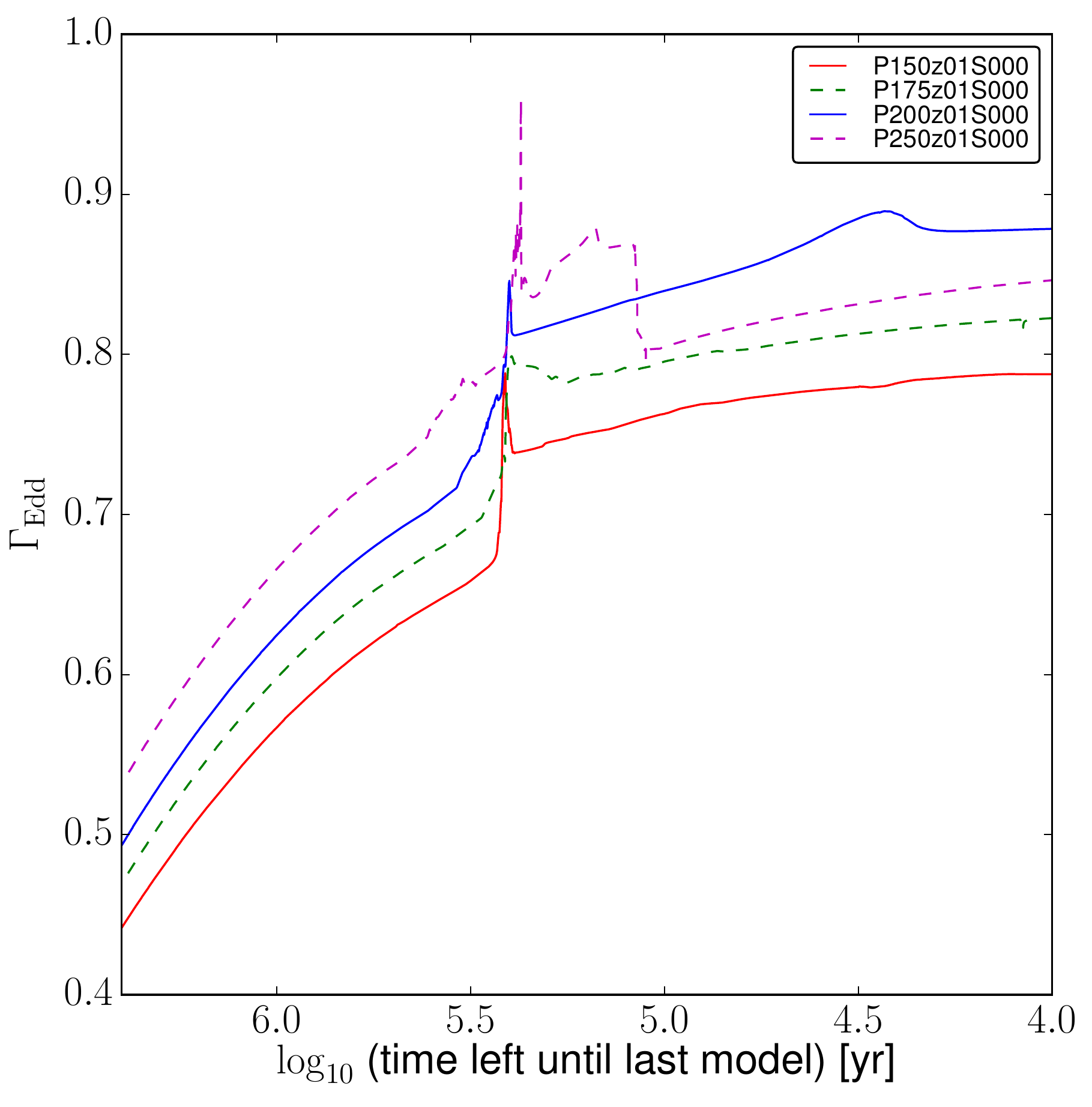}
\caption{{\it Left}: Evolution of the mass loss rate as a function of time left until last model (log scale) for the four VMS models. {\it Right}: Evolution of the Eddington parameter, $\Gamma_{\rm edd}$. These plots are zooming in the early stages where most of the mass loss occurs. Note that the curves do not change significantly between log$_{10}$(time left)$=4$ and the end of the evolution.}
\label{fig:massloss_yr}
\end{figure*}

\begin{figure}
\center
\includegraphics[width=0.5\textwidth]{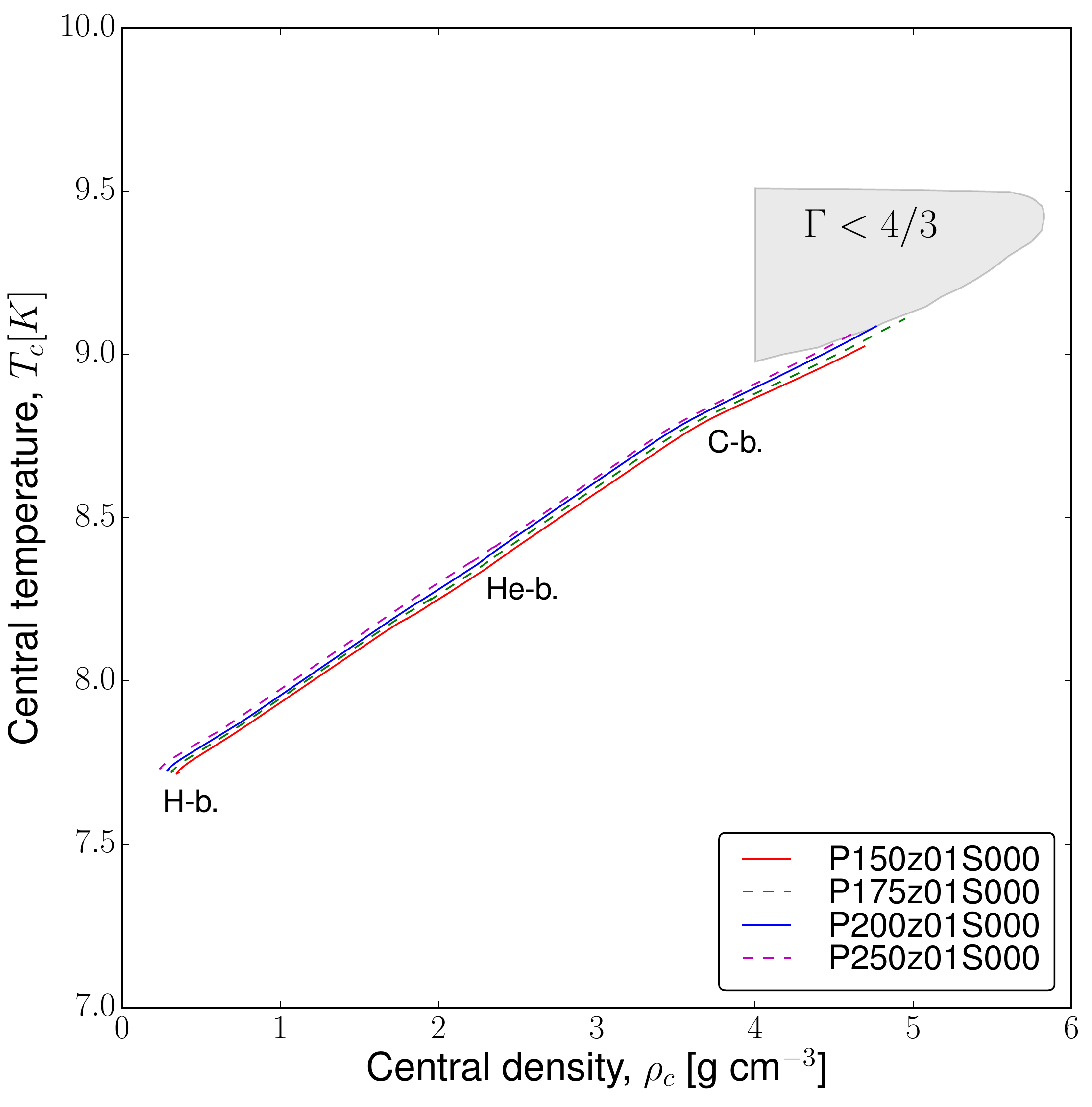}
\caption{Evolution of the central temperature $T_c$ versus central density $\rho_c$ 
for the VMS models. The grey shaded area is the pair-creation instability region 
($\Gamma < 4/3$, where $\Gamma$ is the adiabatic index).  
} \label{fig:tcrhoc}
\end{figure}

The evolution of the models is presented in Figures~\ref{fig:kip_logt} (structure), \ref{fig:hrd} (HRD), \ref{fig:massloss_yr} ({\it left}: mass loss; {\it right}: Eddington parameter),
and \ref{fig:tcrhoc} (central conditions). The models have very large convective cores and are very luminous, which is typical for very massive stars. This leads to very strong mass loss ranging between $10^{-6}$ and $10^{-1}$ solar masses per year. 
The peak in mass loss, seen around log(time left)$\sim 5.4$ in  Figure~\ref{fig:massloss_yr} ({\it left}) corresponds to when the model reach cool parts of the HRD, for which the \citet{deJager88} prescription is used. This is an empirical prescription, which mimics the strong mass loss experienced by luminous blue variable stars. The zigzag pattern (repeated spikes, best seen in model P150 at the end of its evolution) is due to the star getting cooler than $\log (T_\text{eff}) < 3.9$, the mass loss rates switches from the \citet{VN01} to the \citet{deJager88}. This leads to a sharp increase in mass loss, which causes a contraction of the surface back to hotter temperature. Mass loss decreasing by to the \citet{VN01} prescription, the star expands again and the cycle continues creating a zigzag pattern. The very strong mass loss explains why model P250 loses not only most of its hydrogen-rich envelope but also most of its helium envelope, ending its life as a compact CO core. Model P250 is very similar to more metal-rich models as those presented in \citet{Yusof2013}, to which we refer the reader for more details about the evolution of the structure and mass loss in very massive stars. The absence of an extended envelope surrounding the exploding core should lead to a much faster rise to peak luminosity, closer to that of some of the slowest rising SLSNe \citep{Kozyreva2017}.

As the models in this study have a sub-solar metallicity ($Z=0.001$), models P150, P175, and P200 manage to retain a fraction of their hydrogen-rich envelope. They thus are located in cooler parts of the HR-diagram and have larger radii at the end of their evolution than model P250. This is more typical of very low and metal-free models of VMS \citep[see][and references therein]{H07,ES08,YH11,YDL12,CE12,Kozyreva2017}

As expected the fraction of the initial mass lost due to stellar winds increases with initial mass due to the dependence of the mass loss on luminosity. Mass loss prescriptions for VMS are still uncertain though and VMS get close to the Eddington limit towards the end of their evolution. \citet{Grafener11} suggested enhanced mass-loss rates \citep[with respect to][]{VN01} for stars with high Eddington parameters ($\Gamma_{\rm Edd} \geq$ 0.7) that they attribute to the 
Wolf-Rayet stage. In the present work, we did not use an increased mass loss rate close to the Eddington limit. In order to know whether it would have had an impact on the models, we discuss here the proximity of our models to the Eddington limit.
Figure~\ref{fig:massloss_yr} ({\it right}) shows the evolution of the Eddington parameter, $\Gamma_\text{Edd}=L/L_\text{Edd}=\kappa L / (4\pi cGM)$.
The initial values for $\Gamma_\text{Edd}$ range between $0.4 - 0.6$, so well below the Eddington limit, $\Gamma_\text{Edd} = 1$, and
below the limiting value of 0.7 where enhanced
mass-loss rates are expected according to \citet{Grafener11}. As the evolution proceed, however, the Eddington parameter increases to values above 0.7.
Additional mass loss may thus be able to remove the rest of the hydrogen-rich envelope during the late stages, even at very low metallicities. The reader is referred to the recent book on VMS for more detail \citep{VMSbook}.

The evolution of the central conditions is shown in Figure~\ref{fig:tcrhoc}. The models were evolved until at least carbon burning. The end point of the track is the point at which the models were mapped into \verb|FLASH|.

\subsection{Hydrodynamic Simulations in 1D}
\label{subsec:hydro_1d}

\begin{figure}[tbh]
\centering
\begin{tabular}{ll}
\includegraphics[width=0.45\textwidth]{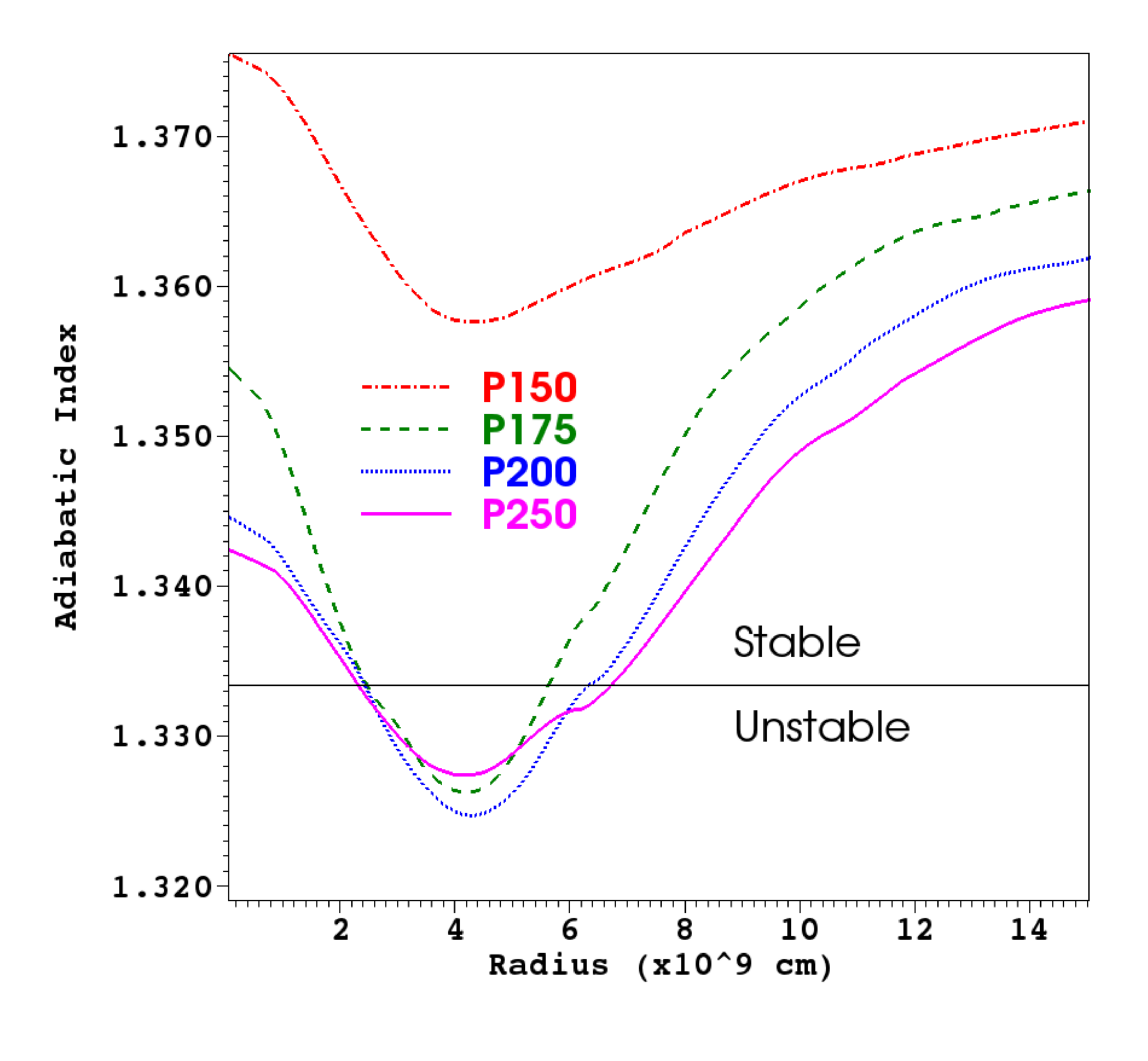}
\end{tabular}
\caption{The adiabatic index in the core of models P150, P175, P200, and P250 at the time of mapping from GENEC to FLASH.}
\label{fig:gamc_prfs}
\end{figure}

The stellar cores from the models described in Table~\ref{tab:models} were mapped into hydrodynamics code \verb|FLASH| (version 4.3) \citep{Fryxell2000,Dubey2009} during core carbon burning, at the end of the evolutionary tracks shown in Fig.~\ref{fig:tcrhoc}. Note that Figure~\ref{fig:tcrhoc} shows the time evolution of the central properties of the models (not the profiles within the star). Figure~\ref{fig:gamc_prfs} shows the profiles of the adiabatic index for the 
\verb|GENEC| model mapped into \verb|FLASH|.
As seen in Figure~\ref{fig:gamc_prfs}, the instability develops first off-center and not in the very center. This explains why the tracks of the central properties (Figure~\ref{fig:tcrhoc}) do not reach the unstable (grey) region. This is due to neutrino cooling being stronger in the very center during the contraction of the core after He-burning. With the exception of P150, the mapping to \verb|FLASH| was done at the point during core carbon burning at which the cores have become slightly unstable due to the PI. P150 constitutes a special case in which the \verb|GENEC| model crashed before reaching the instability (see Figure~\ref{fig:gamc_prfs}). 
The difficulty in evolving this particular model is due to the envelope of the model expanding to large radii following the core contraction at the end of core He-burning, whereas the more massive models (with little or no H-rich envelope) remain compact at the end of their evolution.
The choice of the evolutionary stage for the mapping is critical to generating an explosion with minimum error in stellar evolution. \verb|GENEC| does not include the contribution to the pressure from electron-positron pairs so the mapping must be done before this contribution becomes too great. Conversely, \verb|FLASH| is not a stellar evolution code so the input model must be sufficiently evolved for collapse to occur. We were able to achieve collapse in \verb|FLASH| for models in which the initial pair pressure barely exceeded 1\% of the total pressure in the core. For P150, we had to use a somewhat different method for mapping. We will first explain the method for exploding the three fully evolved models: P175, P200, and P250.

Only the inner cores of radii
$3.334 \times 10^{10}$~cm (P175),
$4.167 \times 10^{10}$~cm (P200),
and
$5.000 \times 10^{10}$~cm (P250) respectively, comprising the CO cores plus a small part of the helium shells were mapped into \verb|FLASH|. For brevity, we will call these inner spheres the `cores'. The radii of the `cores' 
($R_{\text{core}}$)   
were chosen so that the ratio
$ R_{\text{core}}/R_{\text{CO}}$ 
was about 1.3. We mapped the initial models using the same scheme as in \cite{Chatzopoulos2014} 

We use the new directionally-unsplit hydrodynamics solver \citep{LeeDeane2009} coupled with the Aprox19 nuclear burning network \citep{Timmes.Swesty:2000} which are both included in \verb|FLASH|. We employ the Helmholtz equation of state \citep{Timmes.Swesty:2000} and the block-structured grid implementation Paramesh4dev (which includes Adaptive Mesh Refinement; AMR). Self-gravity was computed with the new Multipole solver in \verb|FLASH| \citep{Couch2013c}. The grids were set to an initial minimum resolution of $1.3 \times 10^{8}$~cm with the freedom to refine once, in response to high enough density and/or temperature gradients, to $6.5 \times 10^7$~cm. We employ the `reflect' (`diode') boundary condition for the inner (outer) boundary.

The minimum resolution specified is higher than would normally be required for a PISN simulation. Here, however, we had noticed that mass would slowly flow in through the outer zone in \verb|FLASH| during the time leading up to the collapse (even though the `diode' boundary condition is meant to prevent this).
Since we had to evolve the star in \verb|FLASH| for a considerable amount of time before the collapse occurred, a high minimum resolution was required so that, by the time collapse occurred, only a few solar masses of material had been added to the simulation. The refinement criteria were set to half of their default values in \verb|FLASH| which allowed the maximum refinement level to be achieved in the inner cores during the explosive burning phases. This was important for determining the energetics and nucleosynthetic yields of the explosion. During explosive burning, the shock wave is launched from just outside the exploding core. The simulation is halted just before the shock wave exits the domain (at this point nuclear burning rates are negligible).

To explode P150, we used the same resolution and refinement settings as we did for the rest of the models. The difference is an extra step in between the stellar evolution and the explosion simulation where we use \verb|FLASH|, but with a different hydrodynamics implementation, to prime the model for collapse. We took the final P150 time step from the stellar evolution (shown in Figure~\ref{fig:gamc_prfs}) and mapped the inner 
$1.67 \times 10^{11}$~cm (including the carbon-oxygen core plus part of the helium shell) into \verb|FLASH| to evolve towards the instability using the directionally-split hydrodynamics solver. This solver was better suited for the slow initial approach to the PI. We then re-mapped the unstable core into \verb|FLASH|, this time evolving with the new unsplit hydrodynamics solver as for the other three models to follow the collapse and explosion. 
If the transition is made too early the core will not collapse. If the transition is made too close to collapse the small numerical effects from switching solvers do not have time to dissipate.
Thus, we transition at the earliest time that yields a collapse and explosion. This corresponds to a time when the minimum adiabatic index in the core is 1.302. The subsequent simulation then takes 650.1~s to reach maximum compression, allowing enough time for the core to relax out the numerical effects from the transition. With the additional mapping required for model P150,  the uncertainties are more difficult to quantify. However, we stress that the carbon-oxygen core, the mass of which is an excellent predictor of the explosion properties, was fully formed during the stellar evolution with \verb|GENEC|. In addition, the collapse and explosion phases were fully simulated using the unsplit hydrodynamics solver, which is consistent with the three more massive models.

\subsection{Hydrodynamic Simulations in 2D and 3D}
\label{subsec:multi-D}

Simulations in 1D, 2D, and 3D with each of the four models were done beginning from profiles taken from the 1D simulations described in Section \ref{subsec:hydro_1d} just before collapse (about 20s prior to maximum compression). We used the recommended geometry settings in \verb|FLASH| for each dimensionality: ``spherical'' for 1D, ``cylindrical'' for 2D, and ``Cartesian'' geometry for 3D. Since we were mapping a spherical grid onto non-spherical geometries in the 2D and 3D cases, a slightly smaller domain was evolved (a cube (3D) or half of a square (2D) that could fit inside a sphere (3D) or semicircle (2D) of radius equal to the domain size of the 1D simulation). It was important that the original 1D simulations included a small part of the envelope in their `cores' so that the domains in 2D and 3D could still contain the entire CO cores. The dimensions of our new grids are
in 1D: $0<r<\text{xmax}$ (1D), 
in 2D: $0<r<\text{xmax}$, $-\text{xmax}<z<\text{xmax}$,
and in 3D: $-\text{xmax}<x,y,z<\text{xmax}$,
where for each model xmax had to be less than $1/\sqrt{3}$ ($1/\sqrt{2}$) times the radius of the 1D simulation so that the initial state of the 3D cube (2D half square) could be completely specified by the spherical input model. For P150, P175, P200, and P250 xmax was set to $2.500 \times 10^{10}$~cm,
$1.670 \times 10^{10}$~cm, $2.083 \times 10^{10}$~cm, and $2.500 \times 10^{10}$~cm, respectively. 

We elected to use a fixed grid for the multidimensional simulations after noticing that derefinement led to spurious mixing at shell interfaces. This fixed grid consisted of nested cubes (3D) or half squares (2D) of different, but fixed, resolution settings. In 3D, the innermost cube (with edge length of $1.0 \times 10^{10}$~cm) had the maximum resolution used. This innermost cube was surrounded by a second, larger cube (edge length of $1.5 \times 10^{10}$~cm) which has the next lower refinement level (one factor of two) in the volume not occupied by the innermost cube. The second cube is again surrounded by another (third) cube (edge length of xmax) which has again the next lower refinement level in the volume not occupied by the second cube. In 2D, we used a similar hierarchy of half squares in which the longer edges of the half squares corresponded to the cube edge lengths given above. We use the `diode' boundary condition for all boundaries in 3D and for all boundaries in 2D except the inner boundary for which we use `reflect'. We ran each simulation until the shock reached the edge of the grid which was long after all significant nucleosynthesis had occurred.

To facilitate the comparison of our multidimensional results with our 1D results, we further processed the 3D data by computing angular averages for all the variables. The angular average is a two-step process. First, the state variables are averaged over a `block' ($4^n$ cells, where $n$ is the number of dimensions). Then, we perform another mass-weighted average on to a coarser grid in which the bin widths are $1.5 \times 10^{8}$~cm. During this second average, the $5^{th}$ and $95^{th}$ percentile values for each variable in each bin are also computed. This last step allows us to see how wide the range of values for a particular variable can be within a single radial bin and is used for the shaded regions in Figure \ref{fig:2dv3d} in Section \ref{subsec:exp_in_multid}.

\subsection{Re-appending the Envelope}
\label{subsec:reappend}

For the purpose of comparing to current and future observations, we computed synthetic LCs for all our models. 
For accurate light curves, it was necessary to include the entire star in the simulations. On the timescale of our hydrodynamic simulations (several $\times 10^{4}$~s) we did not expect the envelope to change during the collapse of the core, so we re-appended the pristine envelope from the stellar evolution simulation to the exploding `core' for each model. During the hydrodynamic simulation, the outer edge of the `core' decreased in density and temperature causing a discontinuity between the final `core' profile and the envelope profile. However, we needed to join the two profiles in a smooth manner while still preserving the outer envelope structure which is of the greatest importance to the light curve.

\begin{figure}[tbh]
\centering
\begin{tabular}{ll}
\includegraphics[width=0.45\textwidth]{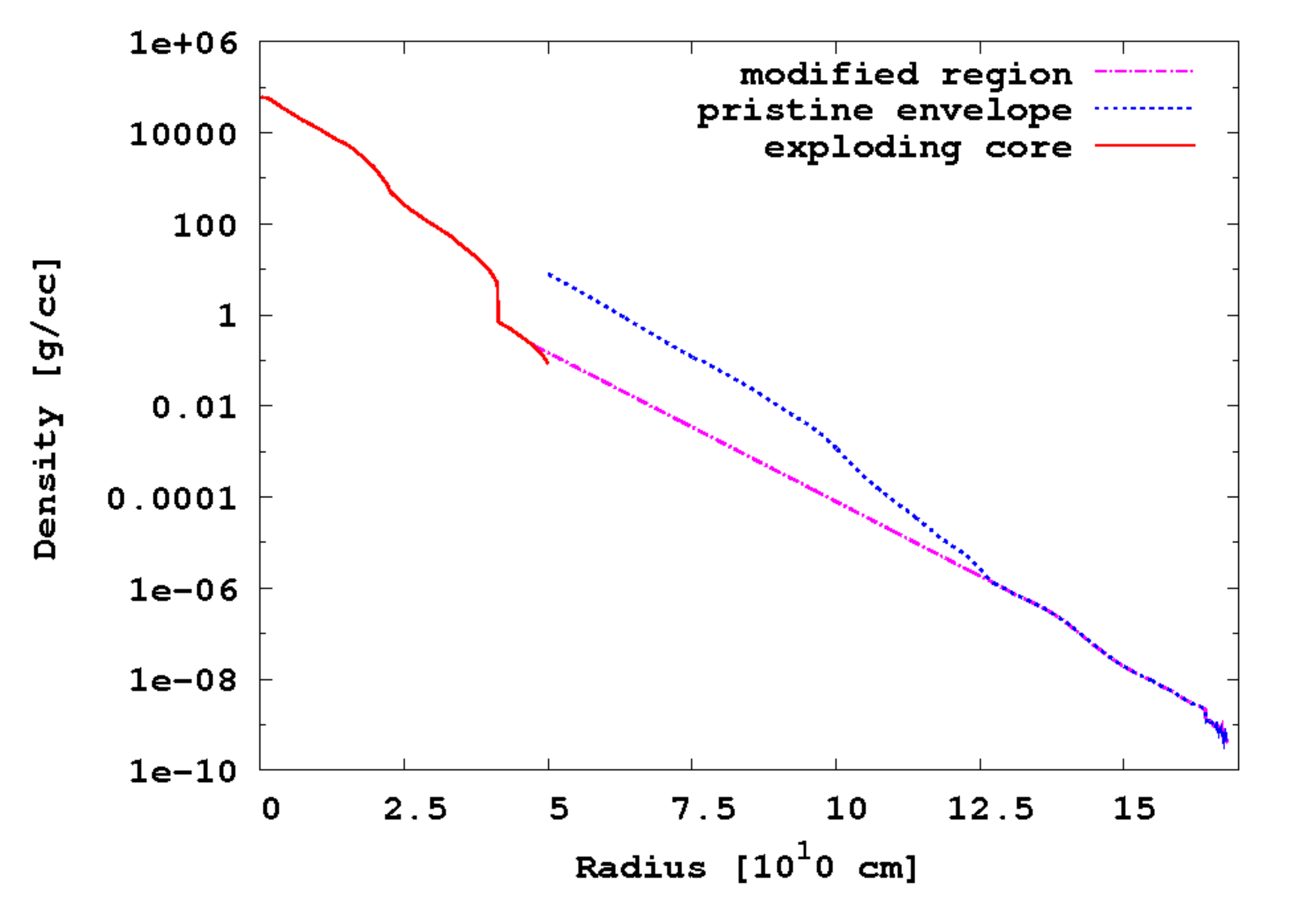}
\end{tabular}
\caption{Density profile including the exploding core (red solid line), the pristine envelope (blue dotted line), and the modified region in between (magenta dashed line).}
\label{fig:env}
\end{figure}

For the 1D simulations, we chose two points between which the density and temperature would be artificially set to follow a straight line connecting the two points in the $\log(\rho)$-radius or $\log(T)$-radius plane. For this modified region, the composition is uniform so the mass fractions are all set according to this uniform composition. One point was chosen to be in front of the shock. The other point was chosen at a point in the envelope and had to fulfill two criteria: (i) the local slope at this point has to be comparable to the slope of the connecting line and (ii) it has to be far enough from the surface so that the structure of the outer envelope was preserved. This method is illustrated using the density profile of P250 as an example in Figure~\ref{fig:env}. The mass and internal energy lost in this process is comparable to but slightly lower than the mass (and accompanied internal energy) that got added through the outer boundary prior to collapse in the \verb|FLASH| simulations. The total mass (energy) changes by less than 3\% (2\%) by this procedure.

In addition to the method described above, we investigated two additional limiting cases. For this test we used a slightly earlier time step than described above because it was necessary in order to apply the first method. Method A involved flattening the density and temperature profiles in the region between the shock and the edge of the core to match the values at the inner edge of the envelope. This method added a small amount of material ($\sim 1 M_{\odot}$) and a small amount of internal energy ahead of the shock. Method B was as described in the paragraph above, except that the exploding core remained untouched. Instead, the density and temperature in the inner region of the envelope were replaced by a straight line connecting the outermost point in the core to the same fixed point in the envelope from before. This method caused a loss of a similarly small amount of material (again, $\sim 1 M_{\odot}$) and internal energy ahead of the shock as method A. For both methods, the resulting profiles were mapped back into \verb|FLASH| for further evolution and then mapped into \verb|STELLA| \citep{Blinnikov2006} using the standard procedure described below and in Section~\ref{sec:stella}) to compute light curves. The resulting light curves were almost identical, indicating that both methods are suitable for this study. Our standard method for re-appending the envelope represents an intermediate method between method A and B, in which matter and energy are removed ahead of the shock (shown by the red line in Figure~\ref{fig:env}). 

In 3D, the simulations were computationally too expensive to include the entire star. In this case,  we conjoined an angular average of the `core' with the pristine envelope using the same method as in the case of the 1D simulations. The only difference is that the angular-averaged `core' profile from the 3D simulation represents a sphere of $2.5 \times 10^{10}$~cm, whereas the pristine envelope begins at $5 \times 10^{10}$~cm due to mapping from a sphere (on a spherical 1D grid to a Cartesian cube in 3D, as described in Section~\ref{subsec:multi-D}). We artificially set the densities and temperatures as described above (linear slope in the $\log{(\rho)}-r$ and $\log(T)-r$ plane, respectively). The velocities in this region are set to zero as during stellar evolution. For the composition we choose the composition of the envelope (which is almost perfectly uniform throughout the envelope). While this step erases information of this intermediate region, it preserves the information from the multi-dimensional simulation of the `core', such as the amount of outward mixing of nickel.

After re-appending the stellar envelope, the entire star was further evolved in \verb|FLASH| before mapping to the radiation-hydrodyanmics code \verb|STELLA| to compute light curves (see Section~\ref{sec:stella}).
For this, we use a maximum resolution of 
$6.5 \times 10^7$~cm and relax the refinement criteria back to their default values. We also allow a lower minimum resolution of $5.2 \times 10^8$~cm (which is lower than in the simulations of the `cores' only) since the low densities at the outer edge of the envelopes effectively remove the problem of mass inflow at the outer boundary of the computational domain. 
As before, we employ the `reflect' ('diode') boundary condition for the inner (outer) boundary.
We choose to map to \verb|STELLA| after the shock has traversed half the radius of the star. The light curve calculations with \verb|STELLA| described in Section~\ref{sec:stella} use these final profiles.

\section{Explosion Properties}
\label{sec:flash}

\subsection{PISN Explosions in 1D}
\label{subsec:1d}

We present explosion properties of all four models considered here in Table \ref{tab:exp_prop}. The Table also includes yields of Ni, Si, and O as calculated within \verb|FLASH|. Between the lowest mass model (P150) and the highest mass model (P250) the nickel yield for the 1D simulations increases by more than four orders of magnitude. This is a consequence of the steep dependence on density and temperature of the associated nuclear reactions together with the dependence of the strength of the collapse on the mass of the CO core. The dependence on CO core mass of the results are in good agreement with previous work \citep{HW02,Kasen2011,Dessart2013,Whalen2013,Whalen2014,Kozyreva2014a,Chatzopoulos2015}. 

Silicon is produced from primarily oxygen (and some carbon). Then, if the temperatures and densities get high enough (as in P200 and P250), nickel is produced in substantial quantities primarily from fusion of silicon. To first order, the explosion converts some number of solar masses of oxygen into silicon and nickel. Thus, the final oxygen masses are substantially lower than their pre-explosion values while both the silicon and nickel yields are substantially higher than their pre-explosion values. Note however that a substantial amount of oxygen remains unburnt in each case and thus constitutes a large fraction of the ejected mass.

\begin{table*}[tbh]
\centering
\caption{1D and 3D Explosion Properties for P150, P175, P200, and P250 at the fiducial resolution ($6.5 \times 10^{7}$~cm).}
\label{tab:exp_prop}
\begin{tabular}{llllllllllllll}
\tableline \tableline
 Model  & \multicolumn{1}{c}{M$_{\text{CO}}$ (M$_{\odot}$)} & \multicolumn{2}{c}{E$_{\text{exp}}$ (B)} & \multicolumn{2}{c}{$\rho_{\text{c}}$ ($10^{6}$~g/cm$^3$)} & \multicolumn{2}{c}{T$_{\text{c}}$ ($10^{9}$~K)} & \multicolumn{2}{c}{M$_{\text{Ni}}$ (M$_{\odot}$)} & \multicolumn{2}{c}{M$_{\text{Si}}$ (M$_{\odot}$)} & \multicolumn{2}{c}{M$_{\text{O}}$ (M$_{\odot}$)} \\
   & & 1D & 3D & 1D & 3D & 1D & 3D & 1D & 3D & 1D & 3D & 1D & 3D \\
\tableline
P150 & 65.7 & 5.58 & 5.68 & 1.42 & 1.39 & 3.41 & 3.39 & $2.86 \times 10^{-3}$ & $2.59 \times 10^{-3}$ & 5.85 & 5.82 & 48.5 & 48.6 \\
P175 & 81.4 & 17.7 & 17.2 & 2.27 & 2.21 & 3.92 & 3.89 & $3.16 \times 10^{-1}$ & $3.00 \times 10^{-1}$ & 12.9 & 12.8 & 51.3 & 51.5 \\
P200 & 100.9 & 49.4 & 48.7 & 3.92 & 3.79 & 4.89 & 4.85 & $1.21 \times 10^{1}$ & $1.16 \times 10^{1}$ & 22.4 & 22.4 & 42.2 & 42.6 \\
P250 & 126.7 & 82.1 & 81.7 & 7.03 & 6.64 & 5.83 & 5.75 & $3.43 \times 10^{1}$ & $3.35 \times 10^{1}$ & 24.5 & 24.6 & 35.9 & 35.7 \\
\tableline
\end{tabular}
\tablecomments{Explosion energies are calculated as the total change in energy during the FLASH simulations plus the initial negative binding energy of the progenitor (so that only exploding models have a positive explosion energy).}
\end{table*}

\begin{figure*}[tbh]
\centering
\includegraphics[width=1.00\textwidth]{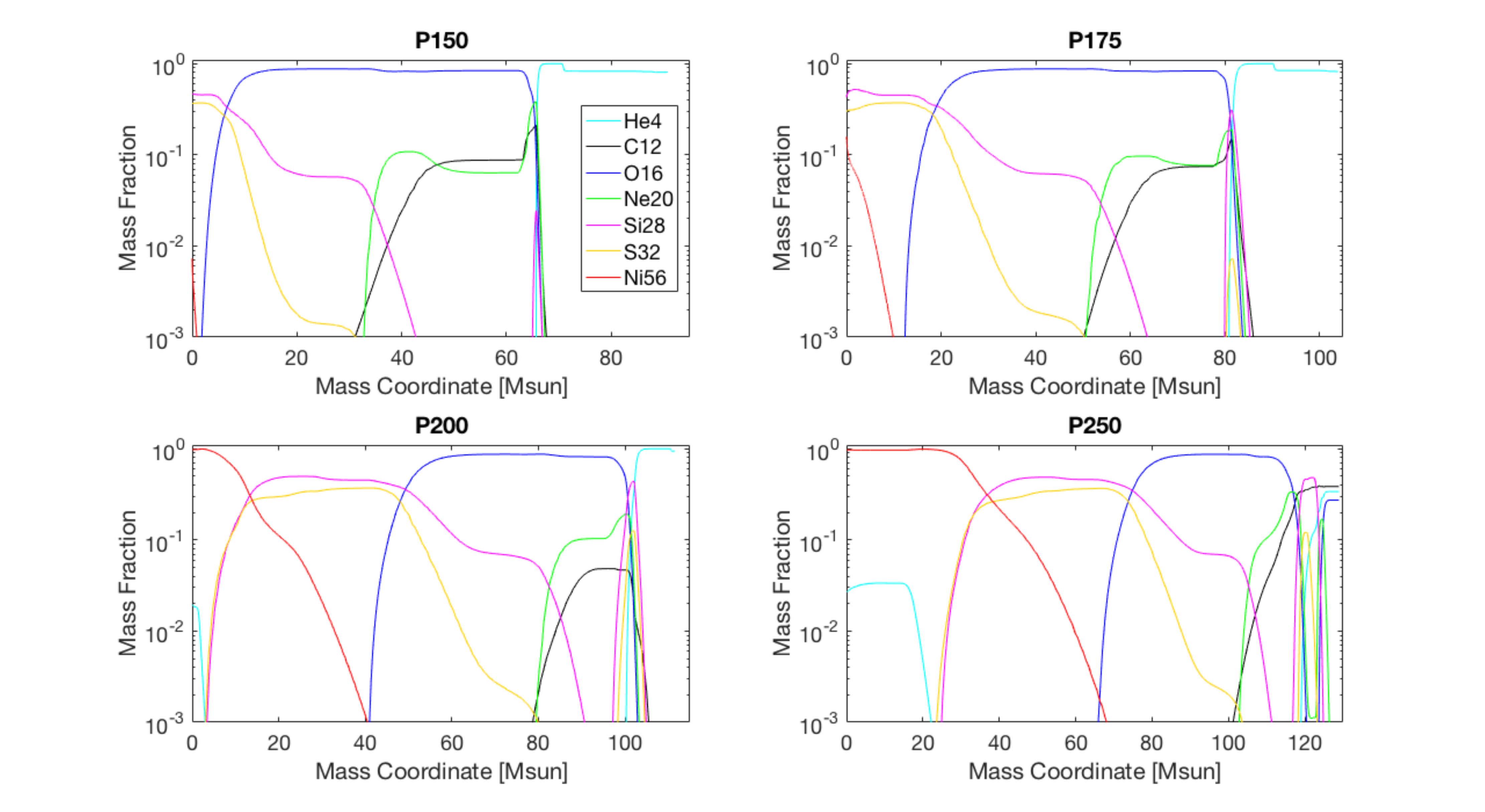}
\caption{Mass Fraction profiles for the 1D simulations of P150 (top left), P175 (top right), P200 (bottom left), and P250 (bottom right) from Table \ref{tab:exp_prop}.}
\label{fig:4panel}
\end{figure*}

Figure~\ref{fig:4panel} shows the compositional profiles of the ejecta after evolution in FLASH is complete. The wide range in nickel yield between the models is evident from the Figure as the higher mass models reach core conditions sufficient for explosive silicon burning in addition to the explosive oxygen burning which is achieved in all the models. Also note how the helium appears in the core for the higher mass models reflecting the dissociation of nickel into helium that occurs due to increasingly high core temperatures. This is the mechanism by which cores of higher mass will collapse directly to black holes. The unburnt oxygen is confined to a shell in between two regions where nucleosynthesis occurred. The region interior to the unburnt oxygen shell experiences the bulk of the nuclear burning, however, there is also a small region (in terms of mass coordinate) outside of this shell in which oxygen captures alpha particles from the inner edge of the stellar envelope. Such burning produces mainly silicon and is triggered by shock heating.

\subsection{PISN Explosions in Multi-D}
\label{subsec:exp_in_multid}

In addition to the simulations in 1D, we have also performed simulations at the same fiducial resolution ($6.5 \times 10^{7}$~cm) in 2D and 3D. For the rest of the paragraph, we focus on the 3D simulations. The 2D simulations yield very similar results for all the properties in Table \ref{tab:exp_prop}. In 3D (and also in 2D) we see a slight decrease in nickel yields (and overall explosion strength) for all models (see Table \ref{tab:exp_prop}) at the fiducial resolution of $6.5 \times 10^{7}$~cm. In 3D, the collapse and explosive burning phases occur with very little asphericity which means that the difference in explosion strengths is likely a resolution effect. The 3D simulations utilized a fixed grid but had the same maximum resolution as the 1D simulations. Although the simulations used the same maximum resolution, the computational cells have different shapes. In 1D the cells are simply radial bins. In 3D, they are cubes whose orientation with respect to the radial direction depends on their position in space. The effective radial resolution is thus spatially dependent and is at best equal to the 1D resolution.  This leads to the expectation that higher resolution is required to achieve convergence in 3D.

To test this, we increased the resolution everywhere in the grid by a factor of two and ran the simulation again in 1D, 2D, and 3D. Because of limitations in computing time, we restrict ourselves to the most massive model, P250. Table \ref{tab:P250_exp_prop} shows the explosion properties in 1D, 2D, and 3D at the new resolution of $3.25 \times 10^{7}$~cm. At this resolution we are fully converged in explosion energy, silicon yield, and nickel yield between the 1D, 2D, and 3D simulations. 
We calculate explosion energies similarly to \citet{Chatzopoulos2015} (see their Table 1):
\begin{equation}
E_{\mathrm{exp}} = E_{\mathrm{tot,f}} - E_{\mathrm{tot,i}} + E_{\mathrm{tot,p}},
\label{eqn}
\end{equation}
where the difference between $E_{\mathrm{tot,f}}$ and $E_{\mathrm{tot,i}}$ is in essence the energy released by nuclear burning and $E_{\mathrm{tot,p}}$ is the initial negative binding energy of the progenitor model. The total energies include contributions from the kinetic, internal, and gravitational potential energies, where the internal energy receives all energy released by nuclear reactions.

\begin{table}[tbh]
\centering
\caption{1D, 2D, and 3D Explosion Properties for P250 at high resolution ($3.25 \times 10^{7}$~cm).}
\label{tab:P250_exp_prop}
\begin{tabular}{llllllll}
\tableline \tableline
   & \multicolumn{1}{l}{E$_{\text{exp}}$} & \multicolumn{1}{l}{$\rho_{\text{c}}$} & \multicolumn{1}{l}{T$_{\text{c}}$} & \multicolumn{1}{l}{M$_{\text{Ni}}$} & \multicolumn{1}{l}{M$_{\text{Si}}$} &
\multicolumn{1}{l}{M$_{\text{O}}$} \\
   & \multicolumn{1}{c}{(B)} & \multicolumn{1}{c}{($10^{6}$~g/cm$^3$)} & \multicolumn{1}{c}{($10^{9}$~K)} & \multicolumn{1}{c}{(M$_{\odot}$)} & \multicolumn{1}{c}{(M$_{\odot}$)} &
\multicolumn{1}{c}{(M$_{\odot}$)} \\
\tableline
1D & 81.9 & 6.90 & 5.80 & 34.0 & 24.5 & 35.6 \\
2D & 81.9 & 6.85 & 5.79 & 34.0 & 24.5 & 35.6 \\
3D & 81.8 & 6.77 & 5.77 & 33.8 & 24.6 & 35.7 \\
\tableline
\end{tabular}
\end{table}

\begin{figure*}
\centering
\includegraphics[width=1\textwidth]{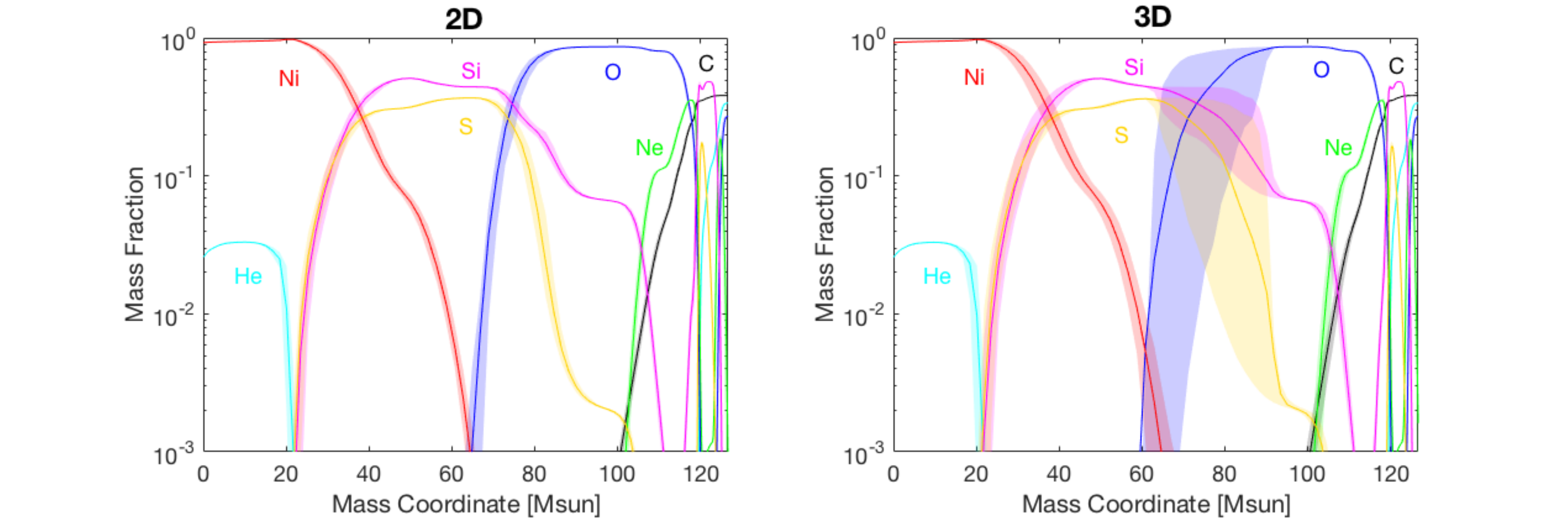}
\caption{Angular-averaged mass fraction profiles for the high resolution ($3.25 \times 10^{7}$~cm) simulations in 2D (left) and 3D (right).}
\label{fig:2dv3d}
\end{figure*}

We now turn our attention to the multidimensional effect of mixing in the supernova ejecta. The mixing is caused by the growth of the Rayleigh-Taylor (RT) instabilities at compositional boundaries. \cite{Chatzopoulos2014} and \cite{Chen2014} have seen this effect in their 2D simulations. 
In our simulations, mixing sets in after maximum compression in the expanding ejecta. The instabilities grow for about 15s at which point the shock reaches the edge of the grid and our simulations are halted.
Figure~\ref{fig:2dv3d} shows the 2D (Left) and 3D (Right) angular averaged profiles at the end of the simulation for model P250 at the increased resolution. The shaded vertical width for each nuclear species is computed for each bin in the angular averaging scheme described in Section \ref{subsec:multi-D} from the distribution of mass fractions involved in the bin average. The lower (upper) edge of the shaded region corresponds to the mass fraction that is greater than 5\% (95\%) of the values occurring in the bin average. Thus, the vertical width can be thought of as a measure of the degree of asphericity for each mass coordinate. Comparing the left and right sides of the figure shows that mixing is stronger in 3D than in 2D. 
This result is not so surprising as it is generally agreed that the RT instability grows faster in 3D than in 2D, at least initially \citep{Kuchugov2014}. Furthermore, \citet{Calder2002} find that this is true in FLASH for single-mode perturbations.

On the right panel of Figure~\ref{fig:2dv3d}, the most apparent feature is the mixing around the Si-O interface. In addition, significant smearing of the interface is apparent in the relatively shallow mass fraction gradients present. We also note weak mixing at the Ni-Si interface, the effect of which is much too small to have an appreciable effect on the LC rise time (see Section~\ref{sec:stella} and Figure~\ref{fig:P2501D3D}). It should be noted that the multidimensional phases of the simulations only followed the evolution until just before the shock exits the grid along the axes (when the shock radius passes 2.5\,$\times$\,10$^{10}$\,cm) while the total radius of the model is 1.67\,$\times$\,10$^{11}$\,cm. Thus, the amount of mixing seen should be regarded as a lower limit to the amount we would expect from the simulation at shock breakout.

\begin{figure*}
\centering
\includegraphics[width=0.45\textwidth]{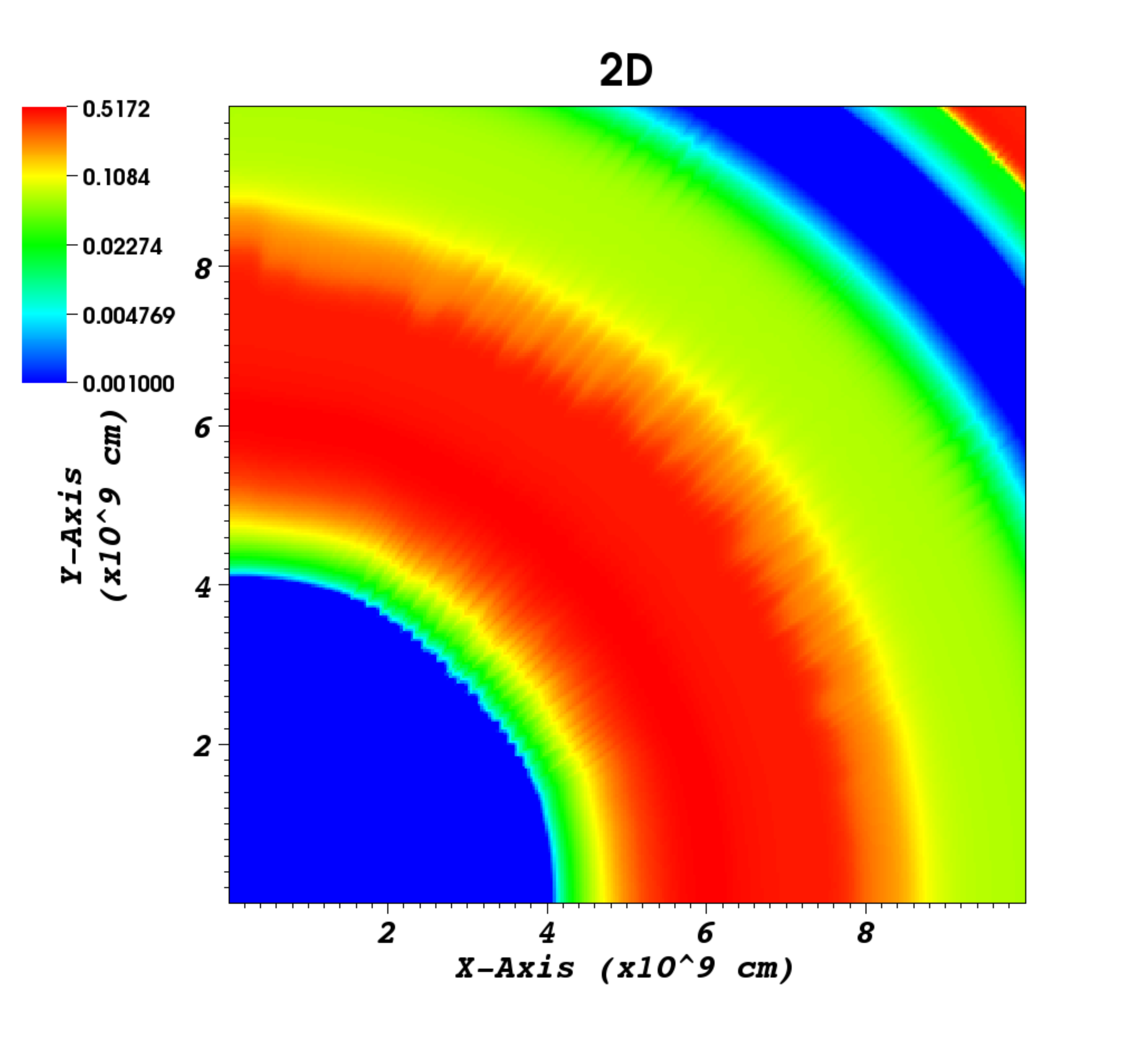}
\includegraphics[width=0.45\textwidth]{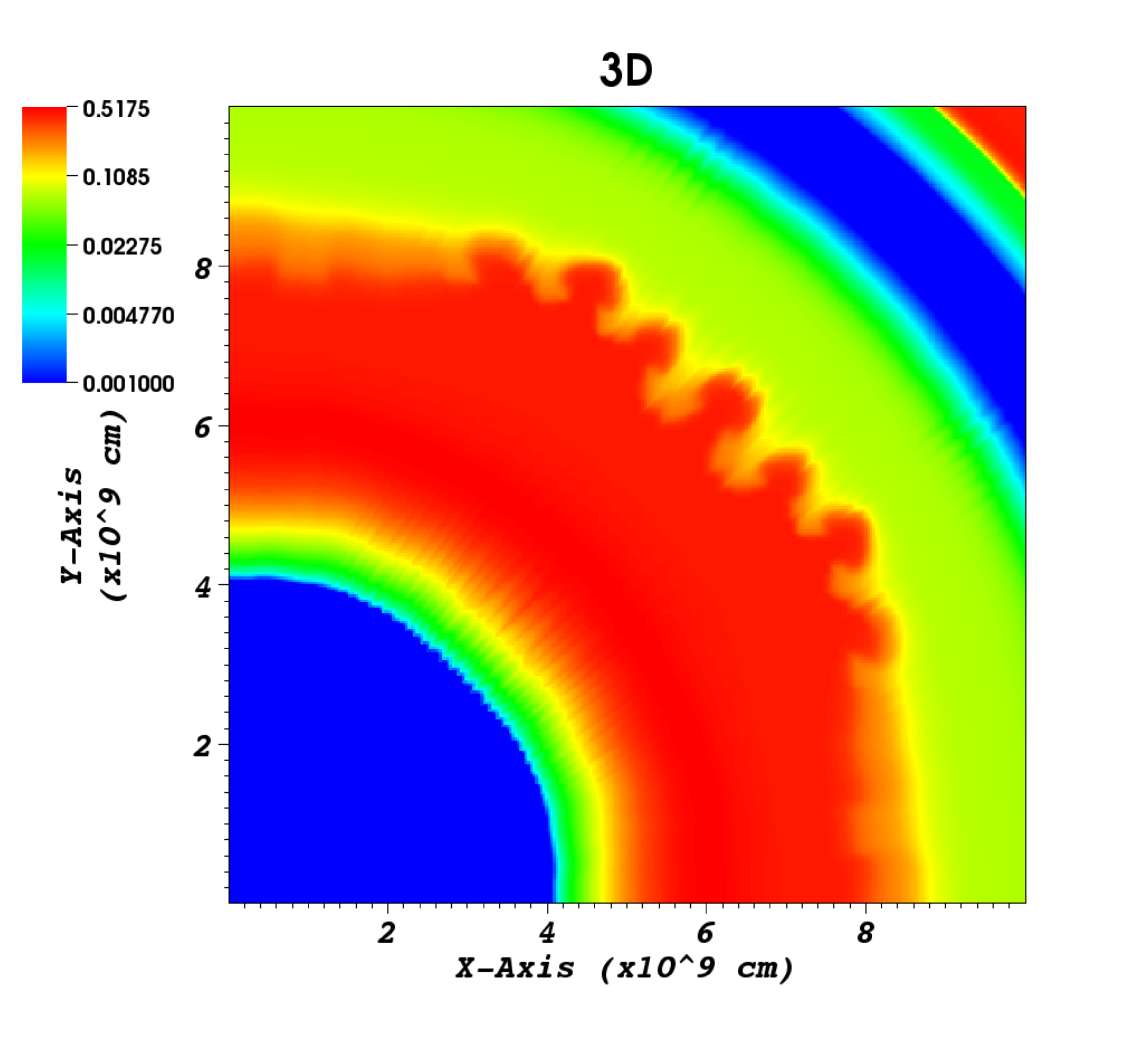}
\caption{Pseudocolor of the $^{28}$Si mass fraction on a log scale for the high resolution ($3.25 \times 10^{7}$~cm) simulations in 2D (left) and 3D (right) centered at the Si-O interface.}
\label{fig:si28}
\end{figure*}

In addition to the strength, the character of the mixing differs between 2D and 3D at the Si-O interface. Figure~\ref{fig:si28} shows the $^{28}$Si mass fraction color coded on a log-scale for the  2D (\emph{left}) and a slice of the 3D (\emph{right}) simulations at the end of the simulation (the same data that was used to generate the $^{28}$Si data for Figure~\ref{fig:2dv3d}). Each panel is centered on the most unstable layer (the Si-O interface). Also visible in the top-right corner of each panel is the inner edge of the outer silicon shell produced via shock burning (see Section~\ref{subsec:1d}). Examining the left panel of Figure~\ref{fig:si28} we note that, in 2D, very thin RT fingers exist on both the inside (the Ni-Si interface) and outside (the Si-O interface) of the Si-rich shell. Focusing now on the right panel, we see that the 3D simulation exhibits similar behavior to the 2D simulation at the Ni-Si interface but very different behavior at the Si-O interface where we see RT plumes whose growth results in much stronger mixing.

\begin{figure}
\centering
\includegraphics[width=0.45\textwidth]{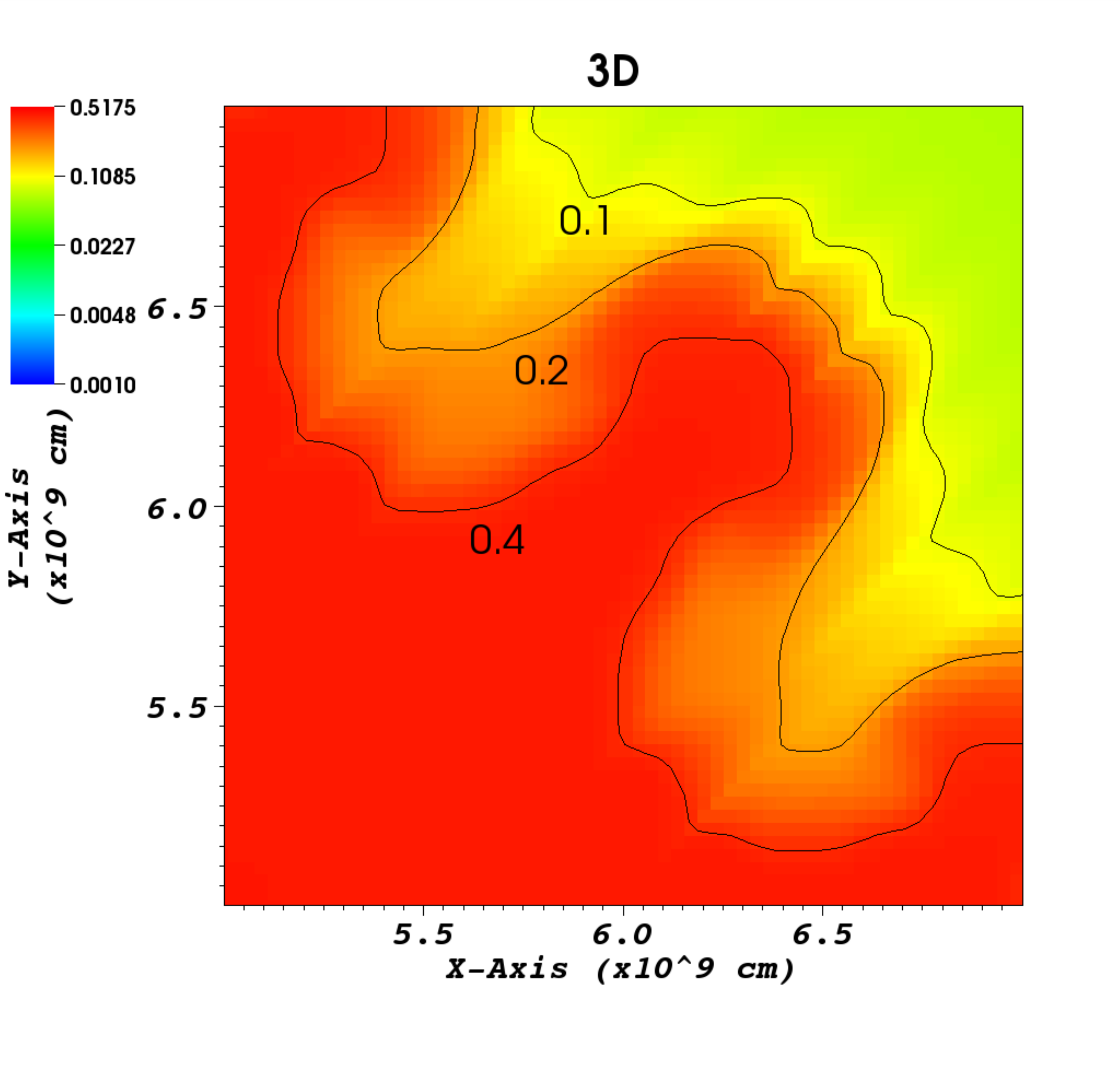}
\caption{25 $\times$ magnification of the RT plume at 45$^{\circ}$ from the right panel of Figure \ref{fig:si28}. Overlayed are contours for three values of the $^{28}$Si mass fraction.}
\label{fig:si28_zoom}
\end{figure}

Also evident from Figure~\ref{fig:si28} is the angular dependence of the RT features. There is a distinct lack of RT plumes directed along the axes. The Cartesian grid has certainly imposed some numerical artifacts on the data. If mixing at this interface occurs in nature, as we expect, there would be no such angular dependence. However, it is unclear if there is an artificial suppression of RT mixing along the axes or if RT mixing is enhanced off axis. Simulations using a spherical geometry may help to shed light on this issue. Additionally, there is evident substructure in the RT plumes. We provide Figure~\ref{fig:si28_zoom} for a closer look at the plume at 45$^{\circ}$ in the right panel of Figure~\ref{fig:si28}. At this scale one can even make out the computational cells. Even finer features may develop in a similar simulation with higher resolution that could further change the character (and perhaps also the strength) of the mixing. We leave this prospect for future work.

\begin{figure*}
\centering
\includegraphics[width=0.45\textwidth]{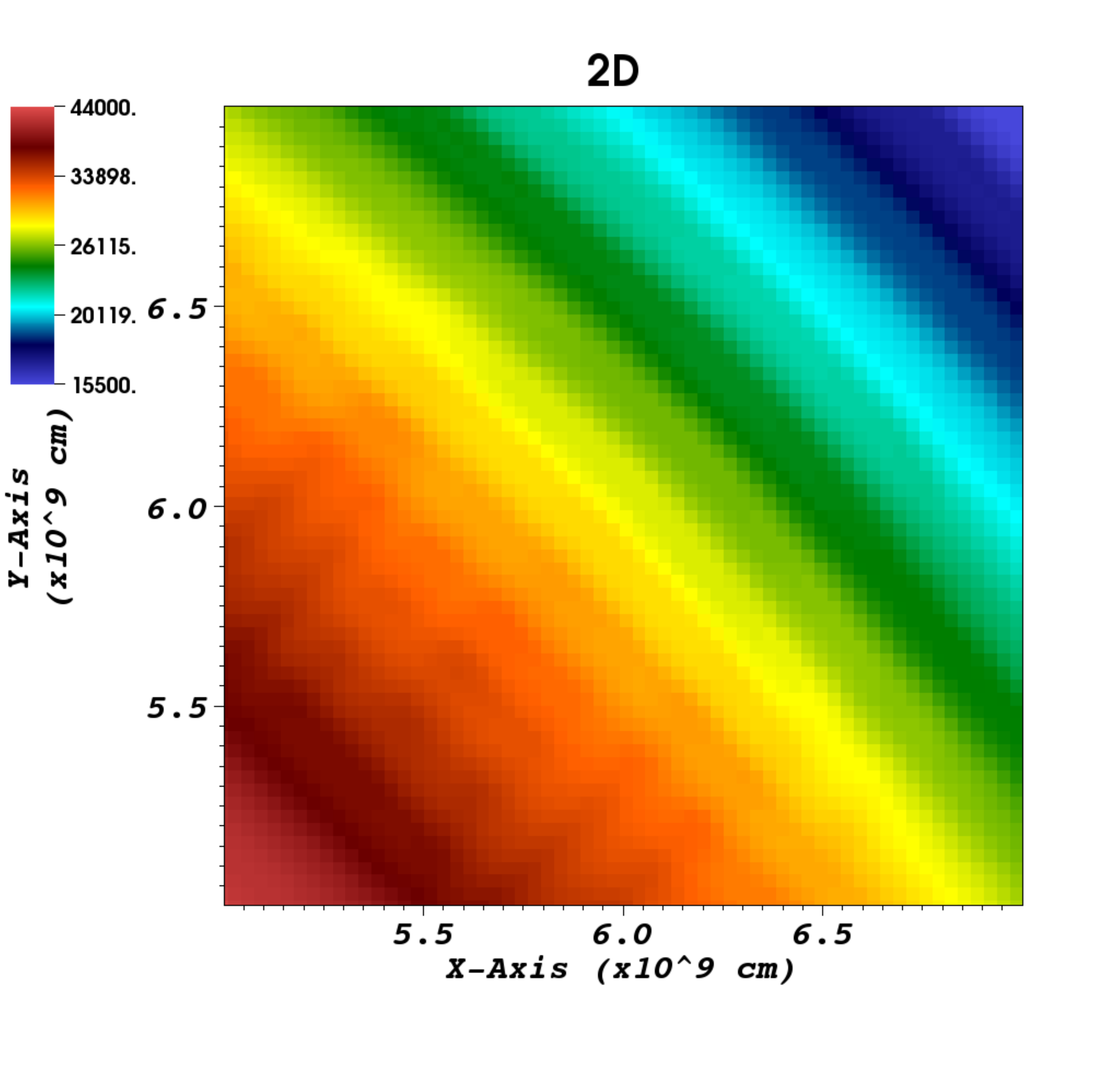}
\includegraphics[width=0.45\textwidth]{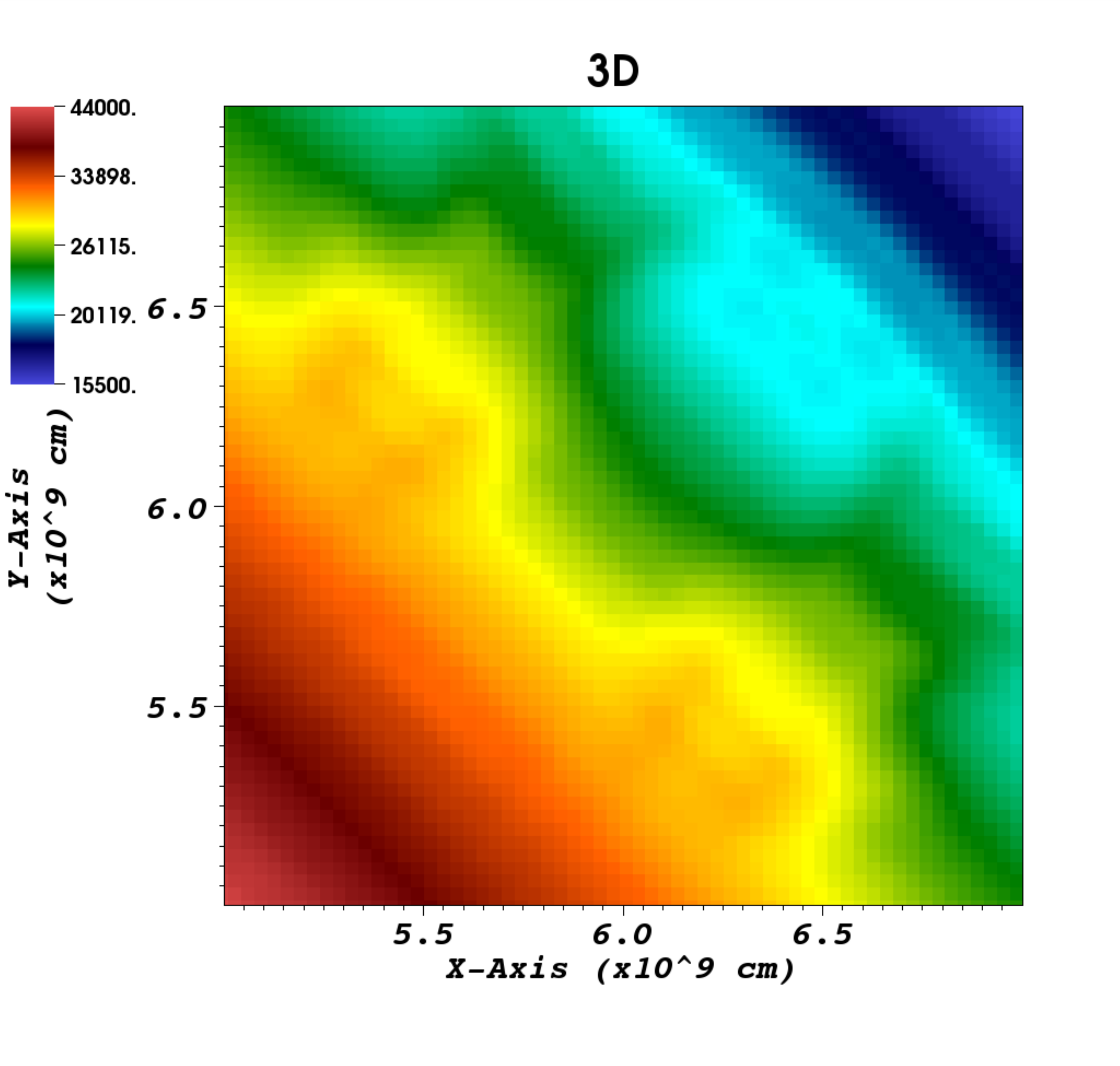}
\caption{Pseudocolor of density on a log-scale for the high resolution ($3.25 \times 10^{7}$~cm) simulations in 2D (left) and 3D (right). Note the smaller scale compared to Figure~\ref{fig:si28}.}
\label{fig:dens}
\end{figure*}

The contrasting character of the RT mixing between 2D and 3D is also evident from the densities at the Si-O interface. Figure~\ref{fig:dens} gives a zoomed-in view (the same view as in Figure~\ref{fig:si28_zoom}) but with density color coded on a log-scale for the 2D (\emph{left}) and the same slice of the 3D (\emph{right}) simulations. We see similar thin fingers and plumes in 2D and 3D, respectively. Note that the underdense regions correspond to the outgoing $^{28}Si$ fingers and plumes. The overdensities of these features are on the order of half of a percent in 2D and 5\% in 3D. Unlike our result that the mixing is stronger in 3D than in 2D, the larger scale of the RT features in 3D compared to 2D is quite surprising and needs further investigation.

 \section{Light Curves}
\label{sec:stella}

We used the 1D radiation-hydrodynamics code \verb|STELLA| (see \cite{Blinnikov1998,Blinnikov2000,Kozyreva2017} for details) to follow the post-explosion evolution of the PISN ejecta and for calculating light curves. For this, we mapped the output from the \verb|FLASH| simulations into \verb|STELLA| when the shock was halfway in the hydrogen-rich envelope in P150/P175 models, and just before shock breakout in P200/P250 models. When velocity gets close to the speed of light in P200 and P250, we cut the \verb|FLASH| profiles at about 30\% of speed of light, because \verb|STELLA| does not include relativistic corrections for the radiative transfer equations. Using this procedure, we calculated bolometric and broad-band light curves for all four models (P150, P175, P200, and P250). We present synthetic LCs from the 1D explosions for all four models. In addition, we show the LC from the fixed-grid, 3D simulation of model P250 using the highest maximum resolution. Since \verb|STELLA| is a 1D code, an angular average of the 3D \verb|FLASH| simulation of model P250 was used for this LC (see Section~\ref{subsec:reappend}).

\begin{figure}
\centering
\includegraphics[width=0.45\textwidth]{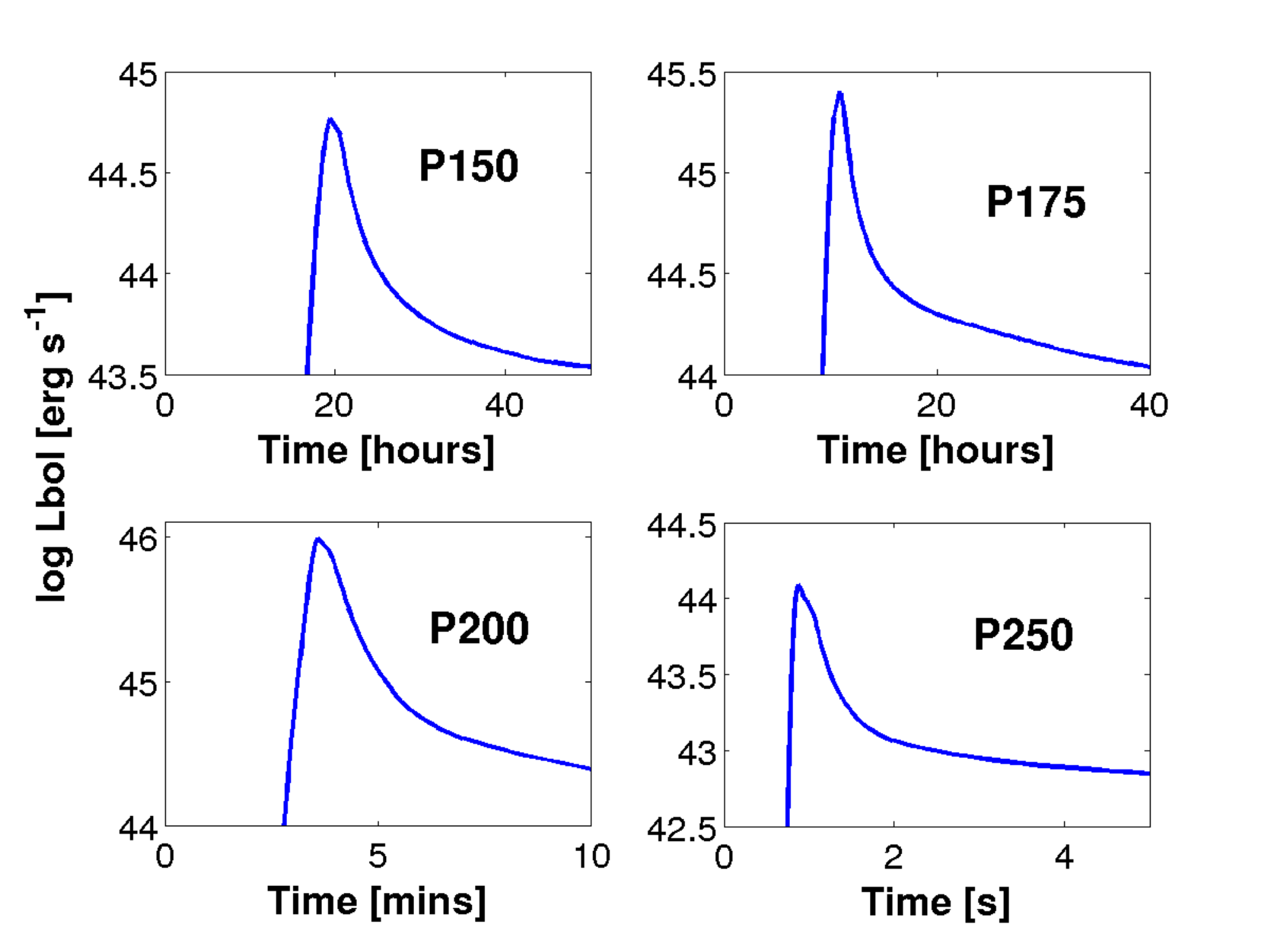}
\caption{Shock breakouts of our PISN models.}
\label{fig:SBO}
\end{figure}

The first light indicating the explosion is radiation emitted as the shock breaks out on the surface of the progenitor. The luminosity at shock breakout depends mostly on the energy of the explosion, the duration depends on the radius of the progenitor. In Figure~\ref{fig:SBO}, we present shock breakouts computed with \verb|STELLA|\footnote{We estimate the duration of the shock breakout event the same way as in \cite{Kasen2011}, i.e. the full width at half-maximum.}. 
The shock breakout lasts for a fraction of a minute for the compact model P250 (0.5~s). At redshift z=10 it will last about 5~s (according to the cosmological time dilation) and will appear in visual or infrared. As for the P200 model, shock breakout lasts 0.7~min (i.e.\ 7~min at z=10). It would be quite challenging to detect such events.  Models P150 and P175 are extended (1107~R$_\odot${} and 1267~R$_\odot${}) and have longer shock breakout durations: 4.2~h{} and 1.7~h{} for P150 and P175, respectively. The color temperature of the shock breakouts are: $7.8 \times 10^5$~K (67~eV) for P150 and P175, $6.5 \times 10^6$~K (560~eV) for P200, and $2.2 \times 10^6$~K (200~eV) for P250.

\begin{figure}[tbh]
\centering
\includegraphics[width=0.48\textwidth]{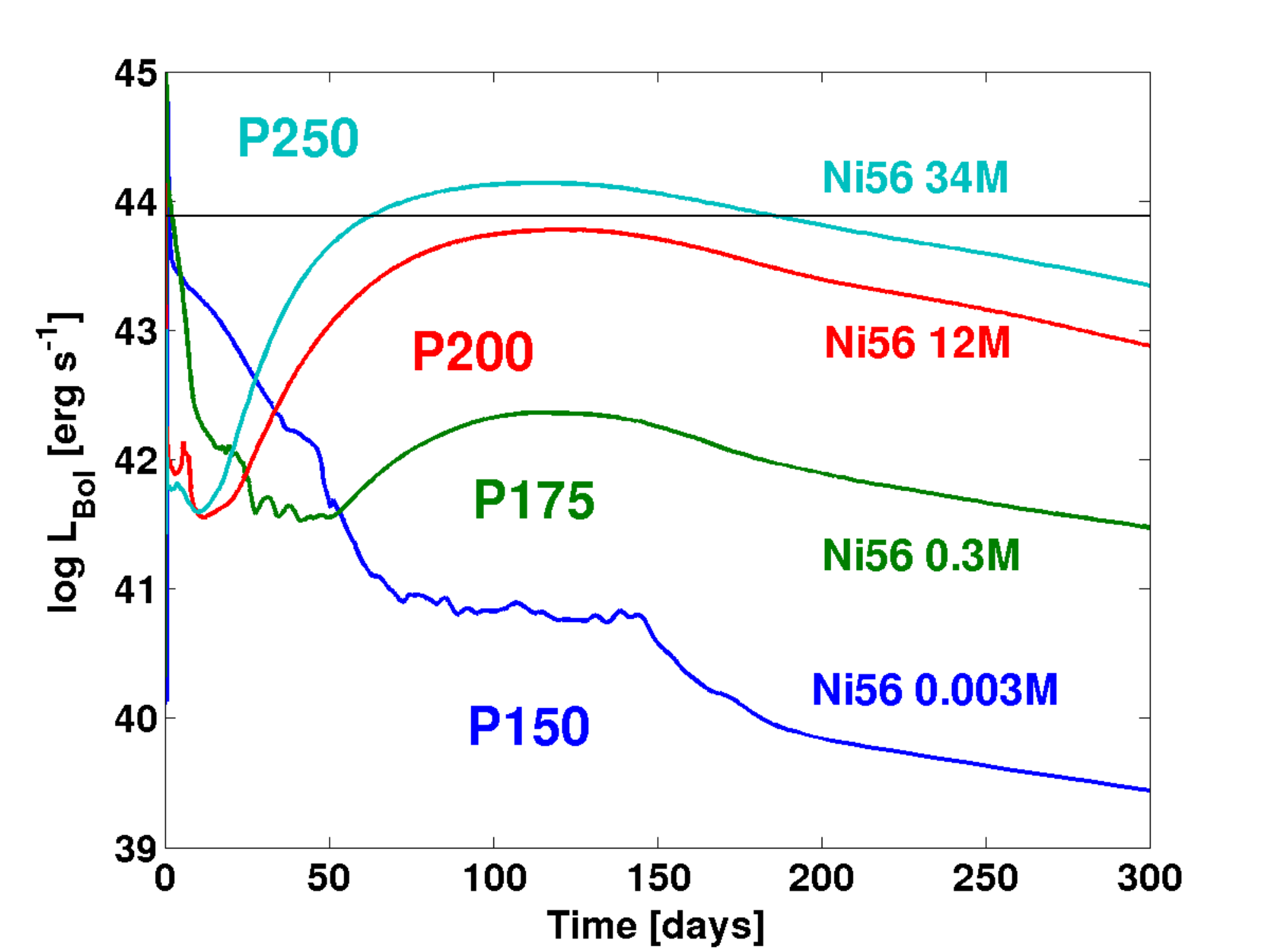}
\caption{Bolometric light curves for P150, P175, P200, and P250 PISN models. The horizontal line at $\log$\,L=43.9\,erg\,s$^{-1}$ corresponds to the SLSN criterion.}
\label{fig:allPISN}
\end{figure}

In Figure~\ref{fig:allPISN}, we present bolometric light curves of all four models\footnote{The light curve data are available via \url{http://www.astro.keele.ac.uk/~kozyreva/LCindex.html}.}. The peak luminosity varies according to the amount of radioactive nickel produced in the PI explosion, i.e.\ from $10^{41}$~erg\,s$^{-1}$ for P150 (M$_{\mathrm{Ni}}=0.003$~M$_\odot${}) to $3\times 10^{44}$~erg\,s$^{-1}$ for P250 (M$_{\mathrm{Ni}}=33.5$~M$_\odot${}). According to the well-established criterion, only P250 is bright enough to be considered a candidate to explain some SLSNe. To generalize, only PISN models with CO-cores above about 110~M$_\odot${}, i.e.\ which produce more than 15~M$_\odot${} of $^{56}$Ni, may result in SLSN-like events. Our P200 model looks like a luminous (but not superluminous) SN, while P175 stays around a luminosity of $10^{42}$~erg\,s$^{-1}$, which is typical for type II SNe. One of the main properties of our PISN models is the width of the light curves. The light curves rise and slowly decay during 200~days, because of high ejecta mass, about 100~M$_\odot${}. The nickel-powered maximum phase resembles the plateau phase of a type IIP SN where the luminosity changes less than 1 magnitude over hundreds of days. Light curves of P175, P200, and P250 peak around day~110, i.e. relatively soon after the explosion compared to the previously published PISN light curves. We refer the reader to \citet{Kozyreva2017} for a discussion of  the reason for the 100-day rise-time, which is shorter by about 50~days than previously published PISN light curves \citep{Kasen2011,Dessart2013}. Our P200 and P250 models are hydrogen-free, therefore, once exploded they appear as hydrogen and helium-free SNe (Type Ic), as photosphere is located deep in the hydrogen and helium-free layers of SN ejecta. 
The P150 light curve is governed by hydrogen recombination during the first 60~days. The photosphere recedes through the extended hydrogen-helium envelope, therefore the luminosity is  relatively high. Later, the P150 light curve flattens because it is supported by oxygen recombination, and luminosity drops as oxygen is located deeper in the ejecta.

\begin{figure*}[tbh]
\centering
\includegraphics[width=0.48\textwidth]{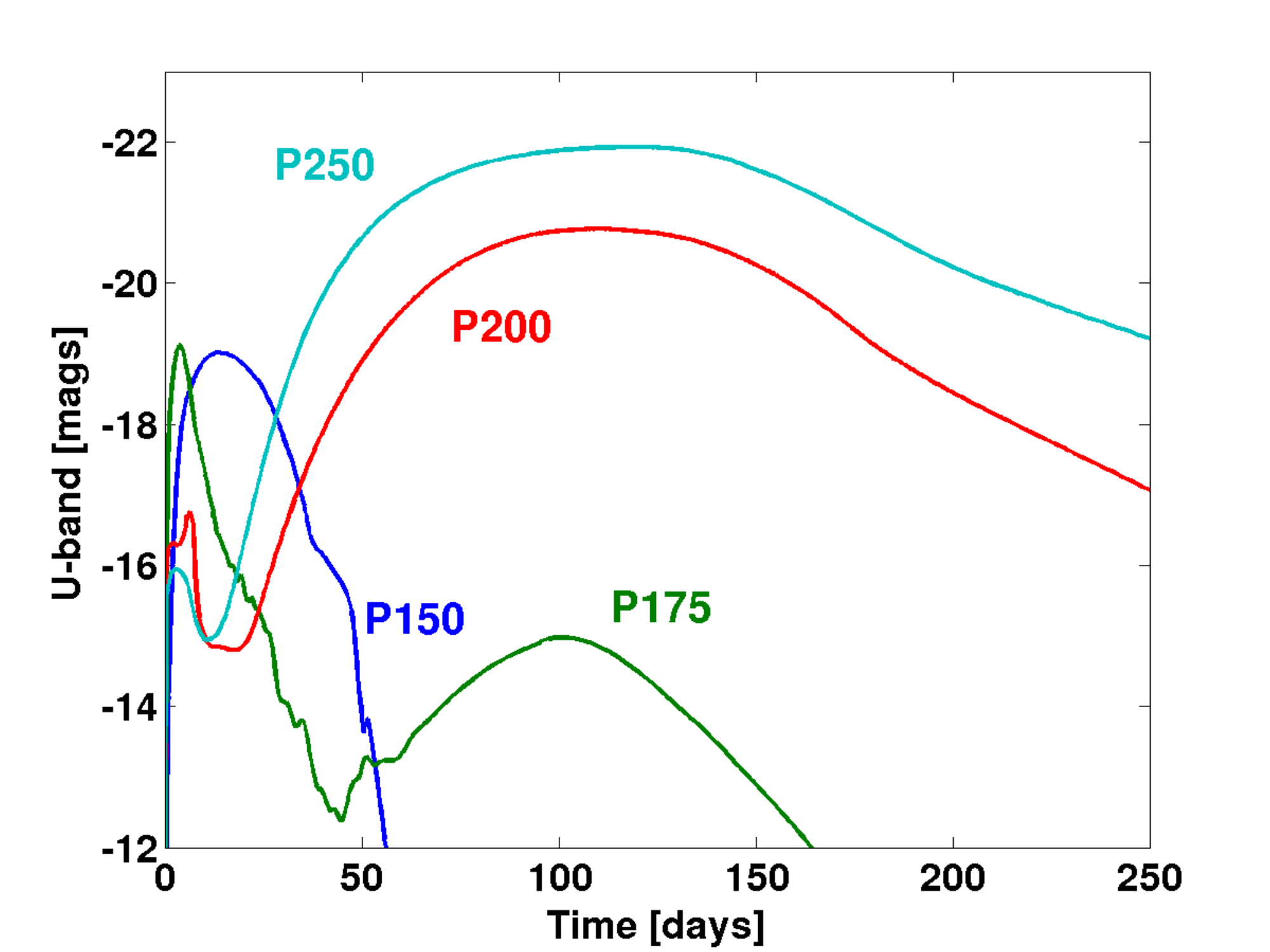}~
\includegraphics[width=0.48\textwidth]{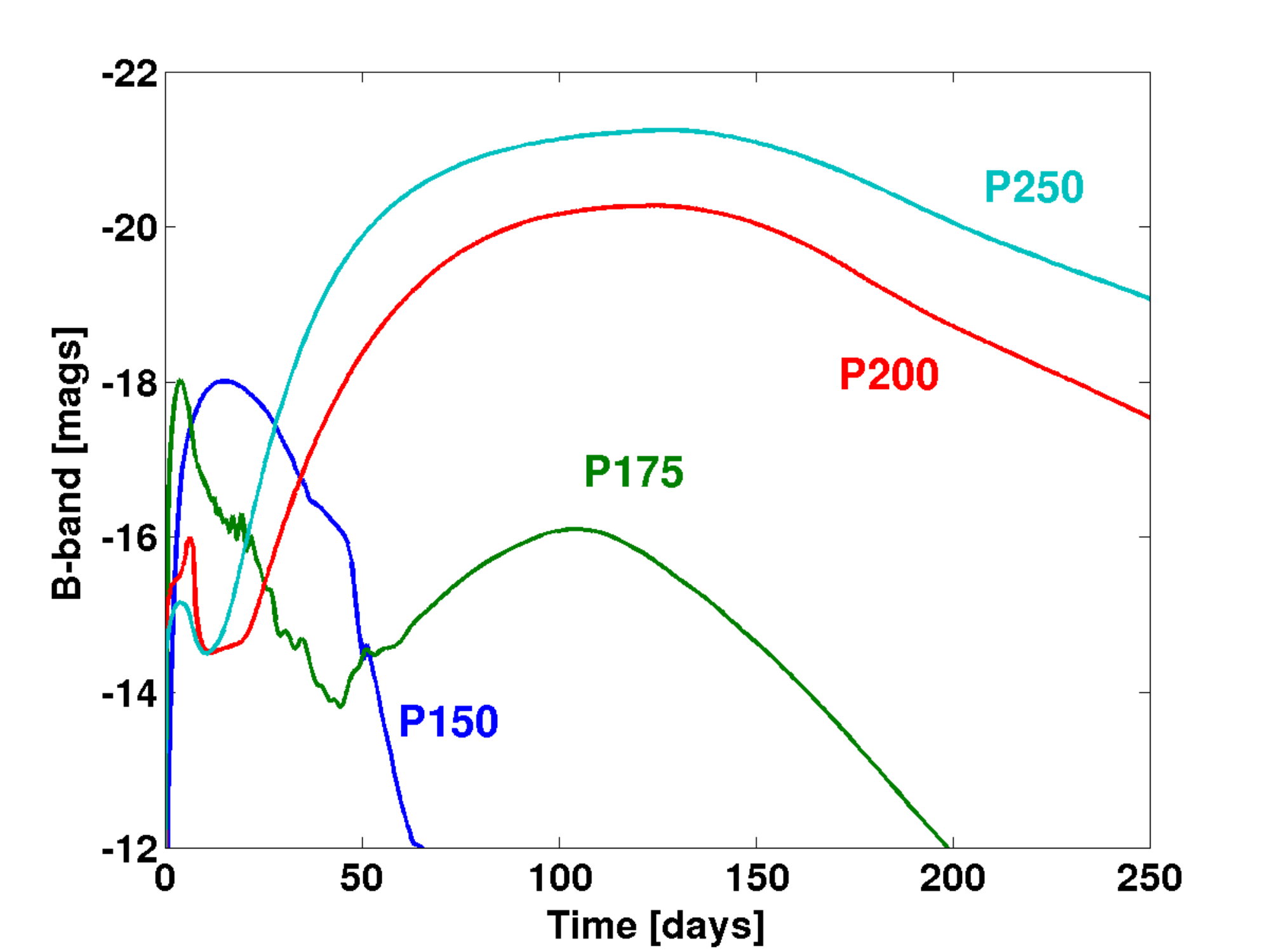}\\
\includegraphics[width=0.48\textwidth]{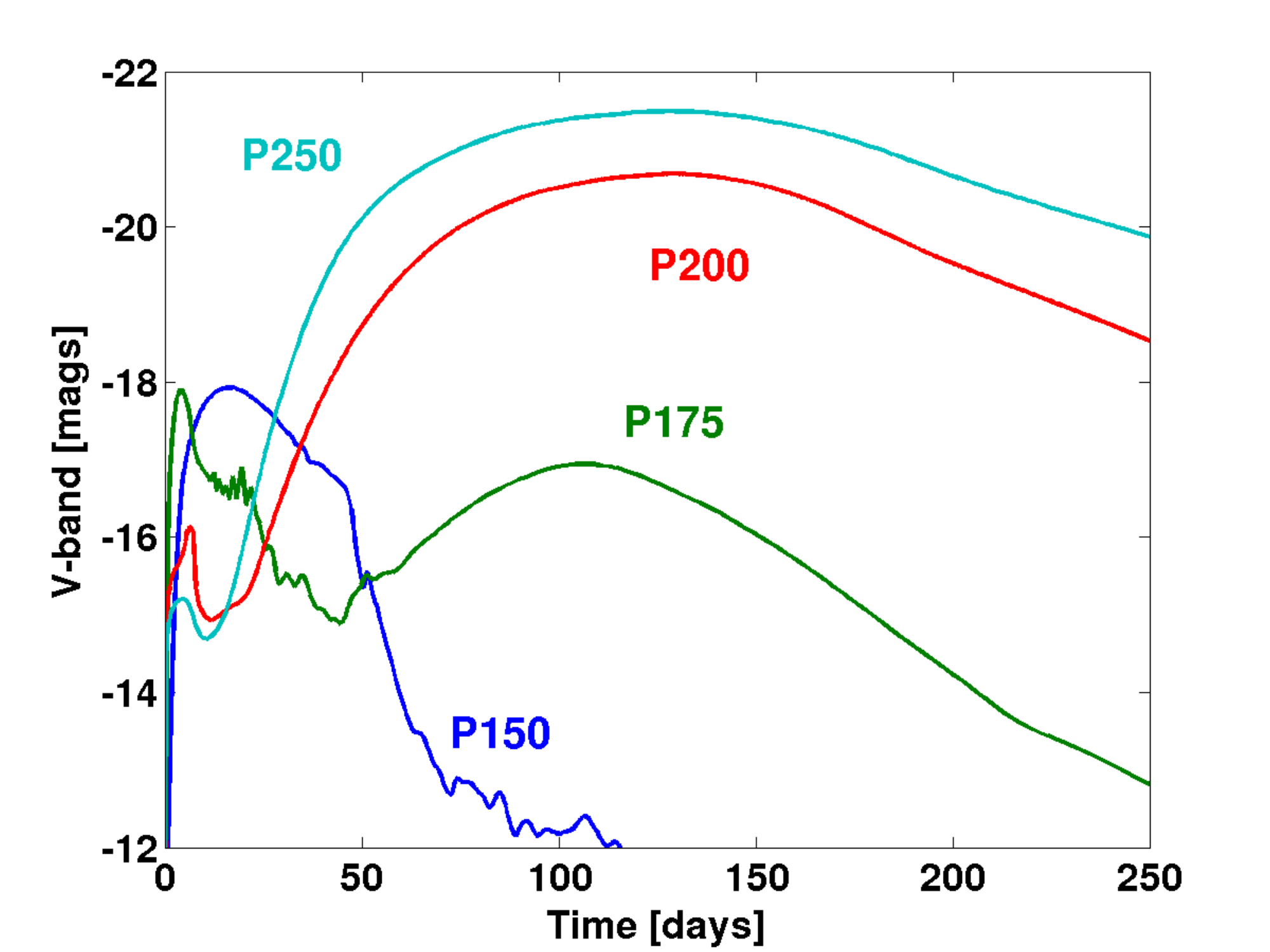}~
\includegraphics[width=0.48\textwidth]{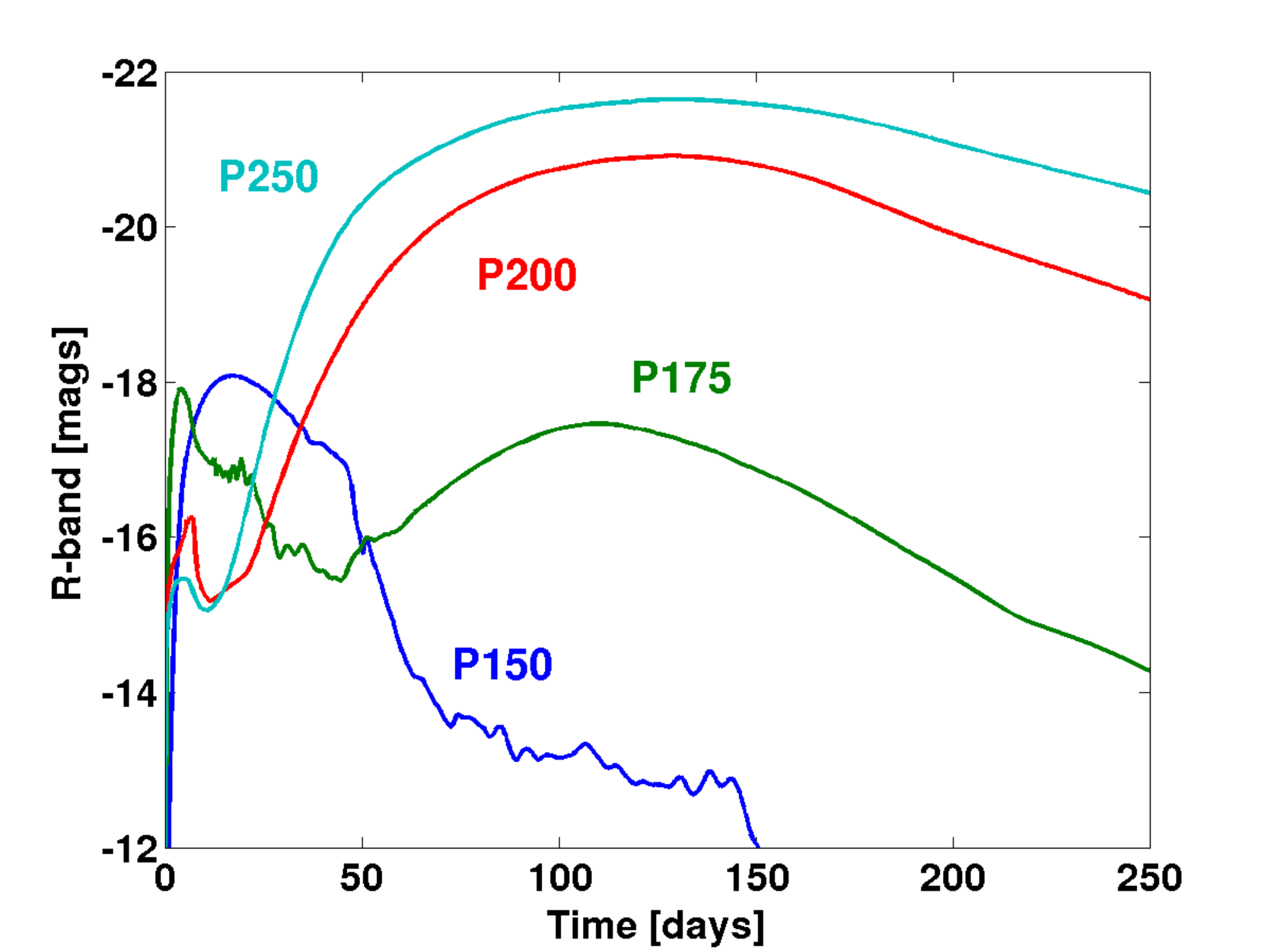}
\caption{Light curves for P150, P175, P200, and P250 PISN models in \emph{U} (top left), \emph{B} (top right), \emph{V} (bottom left), and \emph{R} (bottom right) broad bands.}
\label{fig:bands}
\end{figure*}

In Figure~\ref{fig:bands}, we present light curves for our PISN models in \emph{U}, \emph{B}, \emph{V}, and \emph{R} broad bands. P150 is an outlier compared to other three models, as P150 produces only 0.003~M$_\odot${} of $^{56}$Ni and has no Ni-powered maximum. Instead, P150 has a 50-day earlier maximum supported by a recombination wave receding through the hydrogen-helium envelope.  When compared to the $150$~M$_{\odot}$ red supergiant model 150M by \citet{Kozyreva2014b}, the recombination phase is half as long because our P150 model is relatively less extended (1200~R$_\odot${} for vs.\ 3500~R$_\odot${} for 150M) and contains 3.8~M$_\odot${} of hydrogen versus 4.9~M$_\odot${} in their 150M model (helium yields are 20~M$_\odot${} and 24~M$_\odot${} for P150 and 150M, respectively). The peak magnitude is still high in \emph{U} band reaching --19~mags (--18~mags in \emph{B}, \emph{V}, and \emph{R}), which is similar to plateau magnitudes of bright SNe~IIP \citep{1994AJ....107.1444S}. The other three models (P175, P200, P250) have similarly shaped LCs with a broad peak. P200 and P250 are very bright in all bands, and even brighter in \emph{U} band rather than in \emph{B} and \emph{V} bands. P175 has a short recombination phase similar to P150 (15~days above $-16$~mags in \emph{BVR} bands; 30~days in \emph{U} band) with the peak magnitude $-19$~mags in \emph{U} band and $-18$~mags in other bands. If discovered during the first 30~days P175 will appear as hydrogen-rich type~II SN, while it will be classified as type Ib if discovered at the day~110, i.e. around the main Ni-powered maximum.

\begin{figure}[tbh]
\centering
\includegraphics[width=0.48\textwidth]{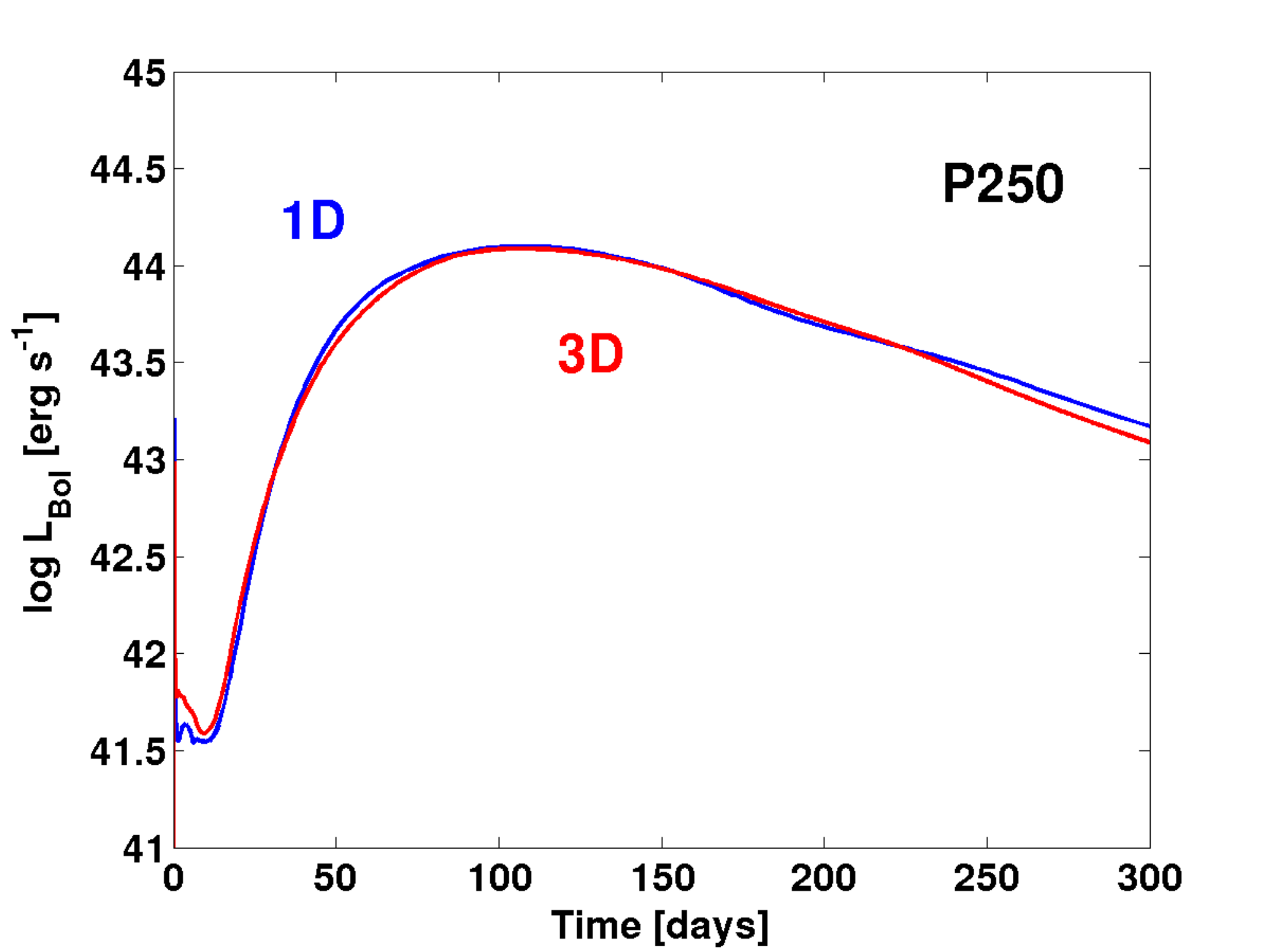}
\caption{Bolometric light curves for P250 PISN models whose explosions are simulated in 1D (blue line) and 3D (red line).}
\label{fig:P2501D3D}
\end{figure}

Finally, in Figure~\ref{fig:P2501D3D}, we compare the bolometric light curves for model P250 from the 1D and 3D explosions. We find that even though there is some degree of mixing at the Ni-Si and \edit2{Si-O} interfaces seen in the right panel of Figure~\ref{fig:2dv3d}, this does not translate to a  significant difference in the bolometric light curves.

\section{Summary}
\label{sec:summary}

PISNe, first theorized in 1967 \citep{Rakavy1967,Barkat1967}, are now a much more intriguing observational prospect than they were initially. It is not known exactly how rare these spectacular explosions are and also whether or not they can occur in the local Universe (many uncertainties still exist in stellar evolution, especially in mass loss rates). However, the physics robustly gives rise to explosions for stars of high enough mass without tuning. The observation of VMS in the local Universe together with simulations of star formation at lower metallicities strongly suggest that PISNe exist. If such an event has already been detected, it is imperative to bridge the gap between modeling and observations. If the first PISN detection is yet to come, models across the mass range at different metallicities and different envelope properties will be useful in identifying such a detection. For these purposes we have further investigated PISN modeling.

We have simulated the life of four VMS models that span the PISN mass range from the main sequence to explosive death and computed the expected LCs. In the process we have used three different astrophysical codes: \verb|GENEC| (stellar evolution), \verb|FLASH| (PISN explosion), and \verb|STELLA| (LC calculation). The mass loss prescriptions used in \verb|GENEC| resulted in progenitor models with very little or no hydrogen at the surface and compact radii compared to published very low or zero metallicity models. Mass loss is the dominant uncertainty in VMS models. The models approach the Eddington limit towards the end of their evolution. This mean that mass loss could be more important in real stars, even at (very) low metallicities. On the other hand, magnetic fields may reduce the efficiency of mass loss \citep{Georgy2017}. STELLA's coupled radiation-hydrodynamics and opacity tables gave relatively fast rise times, helping to alleviate the general discrepancies between PISN model LCs and observations. The versatility of \verb|FLASH| allowed us to simulate the explosions in different dimensionalities and grid settings.

We find that our 1D explosion properties are in good agreement with those of other groups for similar CO core masses. This agreement with other groups, many of which used different stellar evolution and hydrodynamics codes, strengthens the validity of this work as well as previous works which we have compared to.

We extended our study to 2D and 3D to examine the effects of mixing on the ejecta structure and light curves. This extension was strongly motivated by the prospect of shortening the LC rise time by means of outward mixing of $^{56}$Ni \citep{Kozyreva2015}. Such an effect would put our PISN models more in line with the PISN candidates among the already detected SLSNe. Focusing on our most massive model, P250 (the only model that meets the SLSN criterion), we achieved convergence in the explosion properties between all three dimensionalities. However, no significant mixing of $^{56}$Ni was apparent. Indeed, Figure \ref{fig:P2501D3D} shows nearly identical LCs between the 1D and the angle-averaged 3D profiles.

Aside from the lack of a strong $^{56}$Ni mixing affect, we found that mixing was stronger in 3D than in 2D. The compositional interface that experienced the strongest mixing was the Si-O interface. 
Additionally, the character of mixing at this interface differs largely between 2D and 3D. In 2D we see the formation of thin RT fingers and in 3D we see larger-scale plumes. Mixing at this interface will likely have spectral consequences since the spatial distribution in the ejecta directly translates to the velocity distribution. This prospect needs to be explored further as RT growth is strongly resolution dependent.

Finally, we computed synthetic LCs of our four models in 1D, and also for the converged 1D and 3D simulations of P250. Our set of four 1D LCs (shown in Figure \ref{fig:allPISN}) exhibit very luminous shock break-outs and, at peak, span a range in bolometric luminosity greater than  three orders of magnitude, from $10^{41}$~erg\,s$^{-1}$ for P150 to $3\times 10^{44}$~erg\,s$^{-1}$ for P250, according to the amount of radioactive nickel. All exhibit broad plateaus lasting 100--200~days (except P150).

\acknowledgments
Acknowledgments. 
MSG and CF are supported by the United States Department of Energy under an Early CAREER Award (DOE grant no. SC0010263). 
AK was partially supported by the I-Core center of excellence of the CHE-ISF.
NY acknowledges support from University of Malaya Research Grant Programme (UMRG : RP020B-14AFR).
RH acknowledges support from the World Premier International Research Center Initiative (WPI Initiative), MEXT, Japan and from the ``ChETEC'' COST Action (CA16117), supported by COST (European Cooperation in Science and Technology). 
The research leading to these results has received funding from the European Research Council under the European Union's Seventh Framework Programme (FP/2007-2013) / ERC Grant Agreement n. 306901. The \verb|FLASH| and \verb|STELLA| simulations were carried out on the DIRAC Complexity system, operated by the University of Leicester IT Services, which forms part of the STFC DiRAC HPC Facility (www.dirac.ac.uk). This equipment is funded by BIS National EInfrastructure capital grant ST/K000373/1 and STFC DiRAC Operations grant ST/K0003259/1. DiRAC is part of the National EInfrastructure.

\software{GENEC \citep{Ekstrom2012}, FLASH (v4.3; \citet{Fryxell2000} \& \citet{Dubey2009}), STELLA \citep{Blinnikov2006}}

\bibliographystyle{yahapj}
\bibliography{references_push_pisn.bib}

\begin{thebibliography}{}
\providecommand\natexlab[1]{#1}
\providecommand\JournalTitle[1]{#1}

\bibitem[{{Abel} {et~al.}(2002){Abel}, {Bryan}, \& {Norman}}]{Abel2002}
{Abel}, T., {Bryan}, G.~L., \& {Norman}, M.~L. 2002,
  \href{http://dx.doi.org/10.1126/science.295.5552.93}{\JournalTitle{Science},
  295, 93}

\bibitem[{{Barkat} {et~al.}(1967){Barkat}, {Rakavy}, \& {Sack}}]{Barkat1967}
{Barkat}, Z., {Rakavy}, G., \& {Sack}, N. 1967,
  \href{http://dx.doi.org/10.1103/PhysRevLett.18.379}{\JournalTitle{Physical
  Review Letters}, 18, 379}

\bibitem[{{Bennett} {et~al.}(2012){Bennett}, {Hirschi}, {Pignatari}, {Diehl},
  {Fryer}, {Herwig}, {Hungerford}, {Nomoto}, {Rockefeller}, {Timmes}, \&
  {Wiescher}}]{BHP12}
{Bennett}, M.~E., {Hirschi}, R., {Pignatari}, M., {et~al.} 2012,
  \href{http://dx.doi.org/10.1111/j.1365-2966.2012.20193.x}{\JournalTitle{\mnras},
  420, 3047}

\bibitem[{{Blinnikov} {et~al.}(2000){Blinnikov}, {Lundqvist}, {Bartunov},
  {Nomoto}, \& {Iwamoto}}]{Blinnikov2000}
{Blinnikov}, S., {Lundqvist}, P., {Bartunov}, O., {Nomoto}, K., \& {Iwamoto},
  K. 2000, \href{http://dx.doi.org/10.1086/308588}{\JournalTitle{\apj}, 532,
  1132}

\bibitem[{{Blinnikov} {et~al.}(1998){Blinnikov}, {Eastman}, {Bartunov},
  {Popolitov}, \& {Woosley}}]{Blinnikov1998}
{Blinnikov}, S.~I., {Eastman}, R., {Bartunov}, O.~S., {Popolitov}, V.~A., \&
  {Woosley}, S.~E. 1998,
  \href{http://dx.doi.org/10.1086/305375}{\JournalTitle{\apj}, 496, 454}

\bibitem[{{Blinnikov} {et~al.}(2006){Blinnikov}, {R{\"o}pke}, {Sorokina},
  {Gieseler}, {Reinecke}, {Travaglio}, {Hillebrandt}, \&
  {Stritzinger}}]{Blinnikov2006}
{Blinnikov}, S.~I., {R{\"o}pke}, F.~K., {Sorokina}, E.~I., {et~al.} 2006,
  \href{http://dx.doi.org/10.1051/0004-6361:20054594}{\JournalTitle{\aap}, 453,
  229}

\bibitem[{{Bromm} \& {Loeb}(2004)}]{Bromm2004}
{Bromm}, V., \& {Loeb}, A. 2004,
  \href{http://dx.doi.org/10.1016/j.newast.2003.12.006}{\JournalTitle{\na}, 9,
  353}

\bibitem[{{Calder} {et~al.}(2002){Calder}, {Fryxell}, {Plewa}, {Rosner},
  {Dursi}, {Weirs}, {Dupont}, {Robey}, {Kane}, {Remington}, {Drake}, {Dimonte},
  {Zingale}, {Timmes}, {Olson}, {Ricker}, {MacNeice}, \& {Tufo}}]{Calder2002}
{Calder}, A.~C., {Fryxell}, B., {Plewa}, T., {et~al.} 2002,
  \href{http://dx.doi.org/10.1086/342267}{\JournalTitle{\apjs}, 143, 201}

\bibitem[{{Chatzopoulos} {et~al.}(2015){Chatzopoulos}, {van Rossum}, {Craig},
  {Whalen}, {Smidt}, \& {Wiggins}}]{Chatzopoulos2015}
{Chatzopoulos}, E., {van Rossum}, D.~R., {Craig}, W.~J., {et~al.} 2015,
  \href{http://dx.doi.org/10.1088/0004-637X/799/1/18}{\JournalTitle{\apj}, 799,
  18}

\bibitem[{{Chatzopoulos} \& {Wheeler}(2012{\natexlab{a}})}]{Chatzopoulos2012}
{Chatzopoulos}, E., \& {Wheeler}, J.~C. 2012{\natexlab{a}},
  \href{http://dx.doi.org/10.1088/0004-637X/748/1/42}{\JournalTitle{\apj}, 748,
  42}

\bibitem[{{Chatzopoulos} \& {Wheeler}(2012{\natexlab{b}})}]{CE12}
---. 2012{\natexlab{b}},
  \href{http://dx.doi.org/10.1088/0004-637X/748/1/42}{\JournalTitle{\apj}, 748,
  42}

\bibitem[{{Chatzopoulos} {et~al.}(2014){Chatzopoulos}, {Wheeler}, \&
  {Couch}}]{Chatzopoulos2014}
{Chatzopoulos}, E., {Wheeler}, J.~C., \& {Couch}, S.~M. 2014, in American
  Astronomical Society Meeting Abstracts, Vol. 223, American Astronomical
  Society Meeting Abstracts \#223, 335.06

\bibitem[{{Chen} {et~al.}(2014){Chen}, {Heger}, {Woosley}, {Almgren}, \&
  {Whalen}}]{Chen2014}
{Chen}, K.-J., {Heger}, A., {Woosley}, S., {Almgren}, A., \& {Whalen}, D.~J.
  2014,
  \href{http://dx.doi.org/10.1088/0004-637X/792/1/44}{\JournalTitle{\apj}, 792,
  44}

\bibitem[{{Chen} {et~al.}(2015){Chen}, {Smartt}, {Jerkstrand}, {Nicholl},
  {Bresolin}, {Kotak}, {Polshaw}, {Rest}, {Kudritzki}, {Zheng}, {Elias-Rosa},
  {Smith}, {Inserra}, {Wright}, {Kankare}, {Kangas}, \& {Fraser}}]{Chen2015}
{Chen}, T.-W., {Smartt}, S.~J., {Jerkstrand}, A., {et~al.} 2015,
  \href{http://dx.doi.org/10.1093/mnras/stv1360}{\JournalTitle{\mnras}, 452,
  1567}

\bibitem[{{Couch} {et~al.}(2013){Couch}, {Graziani}, \& {Flocke}}]{Couch2013c}
{Couch}, S.~M., {Graziani}, C., \& {Flocke}, N. 2013,
  \href{http://dx.doi.org/10.1088/0004-637X/778/2/181}{\JournalTitle{\apj},
  778, 181}

\bibitem[{{Crowther}(2001)}]{Crowther01}
{Crowther}, P.~A. 2001, in Astrophysics and Space Science Library, Springer,
  ISBN: 0792371046, Vol. 264, The Influence of Binaries on Stellar Population
  Studies, ed. D.~{Vanbeveren}, 215

\bibitem[{{Crowther} {et~al.}(2010{\natexlab{a}}){Crowther}, {Schnurr},
  {Hirschi}, {Yusof}, {Parker}, {Goodwin}, \& {Kassim}}]{PAC10}
{Crowther}, P.~A., {Schnurr}, O., {Hirschi}, R., {et~al.} 2010{\natexlab{a}},
  \href{http://dx.doi.org/10.1111/j.1365-2966.2010.17167.x}{\JournalTitle{\mnras},
  408, 731}

\bibitem[{{Crowther} {et~al.}(2010{\natexlab{b}}){Crowther}, {Schnurr},
  {Hirschi}, {Yusof}, {Parker}, {Goodwin}, \& {Kassim}}]{Crowther2010}
---. 2010{\natexlab{b}},
  \href{http://dx.doi.org/10.1111/j.1365-2966.2010.17167.x}{\JournalTitle{\mnras},
  408, 731}

\bibitem[{{de Jager} {et~al.}(1988){de Jager}, {Nieuwenhuijzen}, \& {van der
  Hucht}}]{deJager88}
{de Jager}, C., {Nieuwenhuijzen}, H., \& {van der Hucht}, K.~A. 1988,
  \JournalTitle{\aaps}, 72, 259

\bibitem[{{Dessart} {et~al.}(2012){Dessart}, {Hillier}, {Waldman}, {Livne}, \&
  {Blondin}}]{Dessart2012}
{Dessart}, L., {Hillier}, D.~J., {Waldman}, R., {Livne}, E., \& {Blondin}, S.
  2012,
  \href{http://dx.doi.org/10.1111/j.1745-3933.2012.01329.x}{\JournalTitle{\mnras},
  426, L76}

\bibitem[{{Dessart} {et~al.}(2013){Dessart}, {Waldman}, {Livne}, {Hillier}, \&
  {Blondin}}]{Dessart2013}
{Dessart}, L., {Waldman}, R., {Livne}, E., {Hillier}, D.~J., \& {Blondin}, S.
  2013, \href{http://dx.doi.org/10.1093/mnras/sts269}{\JournalTitle{\mnras},
  428, 3227}

\bibitem[{{Dubey} {et~al.}(2009){Dubey}, {Reid}, {Weide}, {Antypas},
  {Ganapathy}, {Riley}, {Sheeler}, \& {Siegal}}]{Dubey2009}
{Dubey}, A., {Reid}, L.~B., {Weide}, K., {et~al.} 2009, \JournalTitle{ArXiv
  e-prints}, \href{http://arxiv.org/abs/0903.4875}{{\sffamily arXiv:0903.4875
  [cs.SE]}}

\bibitem[{{Ekstr{\"o}m} {et~al.}(2008){Ekstr{\"o}m}, {Meynet}, {Chiappini},
  {Hirschi}, \& {Maeder}}]{ES08}
{Ekstr{\"o}m}, S., {Meynet}, G., {Chiappini}, C., {Hirschi}, R., \& {Maeder},
  A. 2008,
  \href{http://dx.doi.org/10.1051/0004-6361:200809633}{\JournalTitle{\aap},
  489, 685}

\bibitem[{{Ekstr{\"o}m} {et~al.}(2012){Ekstr{\"o}m}, {Georgy}, {Eggenberger},
  {Meynet}, {Mowlavi}, {Wyttenbach}, {Granada}, {Decressin}, {Hirschi},
  {Frischknecht}, {Charbonnel}, \& {Maeder}}]{Ekstrom2012}
{Ekstr{\"o}m}, S., {Georgy}, C., {Eggenberger}, P., {et~al.} 2012,
  \href{http://dx.doi.org/10.1051/0004-6361/201117751}{\JournalTitle{\aap},
  537, A146}

\bibitem[{{Eldridge} \& {Vink}(2006)}]{EV06}
{Eldridge}, J.~J., \& {Vink}, J.~S. 2006,
  \href{http://dx.doi.org/10.1051/0004-6361:20065001}{\JournalTitle{\aap}, 452,
  295}

\bibitem[{{Fryxell} {et~al.}(2000){Fryxell}, {Olson}, {Ricker}, {Timmes},
  {Zingale}, {Lamb}, {MacNeice}, {Rosner}, {Truran}, \& {Tufo}}]{Fryxell2000}
{Fryxell}, B., {Olson}, K., {Ricker}, P., {et~al.} 2000,
  \href{http://dx.doi.org/10.1086/317361}{\JournalTitle{\apjs}, 131, 273}

\bibitem[{{Gal-Yam}(2012)}]{Gal-Yam2012}
{Gal-Yam}, A. 2012,
  \href{http://dx.doi.org/10.1126/science.1203601}{\JournalTitle{Science}, 337,
  927}

\bibitem[{{Gal-Yam} {et~al.}(2009){Gal-Yam}, {Mazzali}, {Ofek}, {Nugent},
  {Kulkarni}, {Kasliwal}, {Quimby}, {Filippenko}, {Cenko}, {Chornock},
  {Waldman}, {Kasen}, {Sullivan}, {Beshore}, {Drake}, {Thomas}, {Bloom},
  {Poznanski}, {Miller}, {Foley}, {Silverman}, {Arcavi}, {Ellis}, \&
  {Deng}}]{Gal-Yam2009}
{Gal-Yam}, A., {Mazzali}, P., {Ofek}, E.~O., {et~al.} 2009,
  \href{http://dx.doi.org/10.1038/nature08579}{\JournalTitle{\nat}, 462, 624}

\bibitem[{{Georgy} {et~al.}(2017){Georgy}, {Meynet}, {Ekstr{\"o}m}, {Wade},
  {Petit}, {Keszthelyi}, \& {Hirschi}}]{Georgy2017}
{Georgy}, C., {Meynet}, G., {Ekstr{\"o}m}, S., {et~al.} 2017,
  \href{http://dx.doi.org/10.1051/0004-6361/201730401}{\JournalTitle{\aap},
  599, L5}

\bibitem[{{Gr{\"a}fener} \& {Hamann}(2008)}]{GH08}
{Gr{\"a}fener}, G., \& {Hamann}, W.-R. 2008,
  \href{http://dx.doi.org/10.1051/0004-6361:20066176}{\JournalTitle{\aap}, 482,
  945}

\bibitem[{{Gr{\"a}fener} {et~al.}(2011){Gr{\"a}fener}, {Vink}, {de Koter}, \&
  {Langer}}]{Grafener11}
{Gr{\"a}fener}, G., {Vink}, J.~S., {de Koter}, A., \& {Langer}, N. 2011,
  \href{http://dx.doi.org/10.1051/0004-6361/201116701}{\JournalTitle{\aap},
  535, A56}

\bibitem[{{Heger} \& {Woosley}(2002)}]{HW02}
{Heger}, A., \& {Woosley}, S.~E. 2002,
  \href{http://dx.doi.org/10.1086/338487}{\JournalTitle{\apj}, 567, 532}

\bibitem[{{Hirschi}(2007)}]{H07}
{Hirschi}, R. 2007,
  \href{http://dx.doi.org/10.1051/0004-6361:20065356}{\JournalTitle{\aap}, 461,
  571}

\bibitem[{{Inserra} {et~al.}(2017){Inserra}, {Nicholl}, {Chen}, {Jerkstrand},
  {Smartt}, {Kr{\"u}hler}, {Anderson}, {Baltay}, {Della Valle}, {Fraser},
  {Gal-Yam}, {Galbany}, {Kankare}, {Maguire}, {Rabinowitz}, {Smith}, {Valenti},
  \& {Young}}]{2017MNRAS.468.4642I}
{Inserra}, C., {Nicholl}, M., {Chen}, T.-W., {et~al.} 2017,
  \href{http://dx.doi.org/10.1093/mnras/stx834}{\JournalTitle{\mnras}, 468,
  4642}

\bibitem[{{Jerkstrand} {et~al.}(2016){Jerkstrand}, {Smartt}, \&
  {Heger}}]{Jerkstrand2016}
{Jerkstrand}, A., {Smartt}, S.~J., \& {Heger}, A. 2016,
  \href{http://dx.doi.org/10.1093/mnras/stv2369}{\JournalTitle{\mnras}, 455,
  3207}

\bibitem[{{Jerkstrand} {et~al.}(2017){Jerkstrand}, {Smartt}, {Inserra},
  {Nicholl}, {Chen}, {Kr{\"u}hler}, {Sollerman}, {Taubenberger}, {Gal-Yam},
  {Kankare}, {Maguire}, {Fraser}, {Valenti}, {Sullivan}, {Cartier}, \&
  {Young}}]{Jerkstrand2017}
{Jerkstrand}, A., {Smartt}, S.~J., {Inserra}, C., {et~al.} 2017,
  \href{http://dx.doi.org/10.3847/1538-4357/835/1/13}{\JournalTitle{\apj}, 835,
  13}

\bibitem[{{Kasen} {et~al.}(2011){Kasen}, {Woosley}, \& {Heger}}]{Kasen2011}
{Kasen}, D., {Woosley}, S.~E., \& {Heger}, A. 2011,
  \href{http://dx.doi.org/10.1088/0004-637X/734/2/102}{\JournalTitle{\apj},
  734, 102}

\bibitem[{{Kozyreva} \& {Blinnikov}(2015)}]{Kozyreva2015}
{Kozyreva}, A., \& {Blinnikov}, S. 2015,
  \href{http://dx.doi.org/10.1093/mnras/stv2287}{\JournalTitle{\mnras}, 454,
  4357}

\bibitem[{{Kozyreva} {et~al.}(2014{\natexlab{a}}){Kozyreva}, {Blinnikov},
  {Langer}, \& {Yoon}}]{Kozyreva2014b}
{Kozyreva}, A., {Blinnikov}, S., {Langer}, N., \& {Yoon}, S.-C.
  2014{\natexlab{a}},
  \href{http://dx.doi.org/10.1051/0004-6361/201423447}{\JournalTitle{\aap},
  565, A70}

\bibitem[{{Kozyreva} {et~al.}(2014{\natexlab{b}}){Kozyreva}, {Yoon}, \&
  {Langer}}]{Kozyreva2014a}
{Kozyreva}, A., {Yoon}, S.-C., \& {Langer}, N. 2014{\natexlab{b}},
  \href{http://dx.doi.org/10.1051/0004-6361/201423641}{\JournalTitle{\aap},
  566, A146}

\bibitem[{{Kozyreva} {et~al.}(2017){Kozyreva}, {Gilmer}, {Hirschi},
  {Fr{\"o}hlich}, {Blinnikov}, {Wollaeger}, {Noebauer}, {van Rossum}, {Heger},
  {Even}, {Waldman}, {Tolstov}, {Chatzopoulos}, \& {Sorokina}}]{Kozyreva2017}
{Kozyreva}, A., {Gilmer}, M., {Hirschi}, R., {et~al.} 2017,
  \href{http://dx.doi.org/10.1093/mnras/stw2562}{\JournalTitle{\mnras}, 464,
  2854}

\bibitem[{{Krumholz}(2014)}]{Krumholz2014}
{Krumholz}, M.~R. 2014, in Very Massive Stars in the Local Universe, ed. J.~S.
  {Vink} (Springer), 43

\bibitem[{Kuchugov {et~al.}(2014)Kuchugov, Rozanov, \&
  Zmitrenko}]{Kuchugov2014}
Kuchugov, P.~A., Rozanov, V.~B., \& Zmitrenko, N.~V. 2014,
  \href{http://dx.doi.org/10.1134/S1063780X14060038}{\JournalTitle{Plasma
  Physics Reports}, 40, 451}

\bibitem[{{Langer} {et~al.}(2007){Langer}, {Norman}, {de Koter}, {Vink},
  {Cantiello}, \& {Yoon}}]{Langer2007}
{Langer}, N., {Norman}, C.~A., {de Koter}, A., {et~al.} 2007,
  \href{http://dx.doi.org/10.1051/0004-6361:20078482}{\JournalTitle{\aap}, 475,
  L19}

\bibitem[{{Lee} \& {Deane}(2009)}]{LeeDeane2009}
{Lee}, D., \& {Deane}, A.~E. 2009,
  \href{http://dx.doi.org/10.1016/j.jcp.2008.08.026}{\JournalTitle{Journal of
  Computational Physics}, 228, 952}

\bibitem[{{Martins}(2014)}]{Martins2014}
{Martins}, F. 2014, in Very Massive Stars in the Local Universe, ed. J.~S.
  {Vink} (Springer), 9

\bibitem[{{Muijres} {et~al.}(2011){Muijres}, {de Koter}, {Vink}, {Krti{\v
  c}ka}, {Kub{\'a}t}, \& {Langer}}]{muijres11}
{Muijres}, L.~E., {de Koter}, A., {Vink}, J.~S., {et~al.} 2011,
  \href{http://dx.doi.org/10.1051/0004-6361/201014290}{\JournalTitle{\aap},
  526, A32}

\bibitem[{{Nicholl} {et~al.}(2013){Nicholl}, {Smartt}, {Jerkstrand}, {Inserra},
  {McCrum}, {Kotak}, {Fraser}, {Wright}, {Chen}, {Smith}, {Young}, {Sim},
  {Valenti}, {Howell}, {Bresolin}, {Kudritzki}, {Tonry}, {Huber}, {Rest},
  {Pastorello}, {Tomasella}, {Cappellaro}, {Benetti}, {Mattila}, {Kankare},
  {Kangas}, {Leloudas}, {Sollerman}, {Taddia}, {Berger}, {Chornock}, {Narayan},
  {Stubbs}, {Foley}, {Lunnan}, {Soderberg}, {Sanders}, {Milisavljevic},
  {Margutti}, {Kirshner}, {Elias-Rosa}, {Morales-Garoffolo}, {Taubenberger},
  {Botticella}, {Gezari}, {Urata}, {Rodney}, {Riess}, {Scolnic}, {Wood-Vasey},
  {Burgett}, {Chambers}, {Flewelling}, {Magnier}, {Kaiser}, {Metcalfe},
  {Morgan}, {Price}, {Sweeney}, \& {Waters}}]{Nicholl2013}
{Nicholl}, M., {Smartt}, S.~J., {Jerkstrand}, A., {et~al.} 2013,
  \href{http://dx.doi.org/10.1038/nature12569}{\JournalTitle{\nat}, 502, 346}

\bibitem[{{Nugis} \& {Lamers}(2000)}]{NL00}
{Nugis}, T., \& {Lamers}, H.~J.~G.~L.~M. 2000, \JournalTitle{\aap}, 360, 227

\bibitem[{{Quimby} {et~al.}(2012){Quimby}, {Arcavi}, {Sternberg}, {Ben-Ami},
  {Yaron}, {Gal-Yam}, {Graham}, {Cenko}, {Filippenko}, {Perley}, {Cao}, \&
  {Kulkarni}}]{Quimby2012}
{Quimby}, R.~M., {Arcavi}, I., {Sternberg}, A., {et~al.} 2012,
  \JournalTitle{The Astronomer's Telegram}, 4121

\bibitem[{{Rakavy} \& {Shaviv}(1967)}]{Rakavy1967}
{Rakavy}, G., \& {Shaviv}, G. 1967,
  \href{http://dx.doi.org/10.1086/149204}{\JournalTitle{\apj}, 148, 803}

\bibitem[{{Schmidt} {et~al.}(1994){Schmidt}, {Kirshner}, {Eastman}, {Hamuy},
  {Phillips}, {Suntzeff}, {Maza}, {Filippenko}, {Ho}, {Matheson}, {Grashuis},
  {Aviles}, {Kirkpatrick}, {Challis}, {Kuijken}, {Zucker}, {Bolte}, \&
  {Tyson}}]{1994AJ....107.1444S}
{Schmidt}, B.~P., {Kirshner}, R.~P., {Eastman}, R.~G., {et~al.} 1994,
  \href{http://dx.doi.org/10.1086/116957}{\JournalTitle{\aj}, 107, 1444}

\bibitem[{{Stacy} \& {Bromm}(2014)}]{Stacy2014}
{Stacy}, A., \& {Bromm}, V. 2014,
  \href{http://dx.doi.org/10.1088/0004-637X/785/1/73}{\JournalTitle{\apj}, 785,
  73}

\bibitem[{{Stacy} {et~al.}(2016){Stacy}, {Bromm}, \& {Lee}}]{Stacy2016}
{Stacy}, A., {Bromm}, V., \& {Lee}, A.~T. 2016,
  \href{http://dx.doi.org/10.1093/mnras/stw1728}{\JournalTitle{\mnras}, 462,
  1307}

\bibitem[{{Sylvester} {et~al.}(1998){Sylvester}, {Skinner}, \&
  {Barlow}}]{sylvester98}
{Sylvester}, R.~J., {Skinner}, C.~J., \& {Barlow}, M.~J. 1998,
  \href{http://dx.doi.org/10.1046/j.1365-8711.1998.02078.x}{\JournalTitle{\mnras},
  301, 1083}

\bibitem[{{Timmes} \& {Swesty}(2000)}]{Timmes.Swesty:2000}
{Timmes}, F.~X., \& {Swesty}, F.~D. 2000,
  \href{http://dx.doi.org/10.1086/313304}{\JournalTitle{\apjs}, 126, 501}

\bibitem[{{Tornatore} {et~al.}(2007){Tornatore}, {Ferrara}, \&
  {Schneider}}]{Tornatore2007}
{Tornatore}, L., {Ferrara}, A., \& {Schneider}, R. 2007,
  \href{http://dx.doi.org/10.1111/j.1365-2966.2007.12215.x}{\JournalTitle{\mnras},
  382, 945}

\bibitem[{{van Loon} {et~al.}(1999){van Loon}, {Groenewegen}, {de Koter},
  {Trams}, {Waters}, {Zijlstra}, {Whitelock}, \& {Loup}}]{vloon99}
{van Loon}, J.~T., {Groenewegen}, M.~A.~T., {de Koter}, A., {et~al.} 1999,
  \JournalTitle{\aap}, 351, 559

\bibitem[{{Vink}(2014)}]{Vink2014}
{Vink}, J.~S. 2014, in Very Massive Stars in the Local Universe, ed. J.~S.
  {Vink} (Springer), 1

\bibitem[{{Vink}(2015)}]{VMSbook}
{Vink}, J.~S., ed. 2015, Astrophysics and Space Science Library, Vol. 412,
  {Very Massive Stars in the Local Universe} (Springer International Publishing
  Switzerland)

\bibitem[{{Vink} {et~al.}(2001){Vink}, {de Koter}, \& {Lamers}}]{VN01}
{Vink}, J.~S., {de Koter}, A., \& {Lamers}, H.~J.~G.~L.~M. 2001,
  \href{http://dx.doi.org/10.1051/0004-6361:20010127}{\JournalTitle{\aap}, 369,
  574}

\bibitem[{{Whalen} {et~al.}(2013){Whalen}, {Even}, {Frey}, {Smidt}, {Johnson},
  {Lovekin}, {Fryer}, {Stiavelli}, {Holz}, {Heger}, {Woosley}, \&
  {Hungerford}}]{Whalen2013}
{Whalen}, D.~J., {Even}, W., {Frey}, L.~H., {et~al.} 2013,
  \href{http://dx.doi.org/10.1088/0004-637X/777/2/110}{\JournalTitle{\apj},
  777, 110}

\bibitem[{{Whalen} {et~al.}(2014){Whalen}, {Smidt}, {Heger}, {Hirschi},
  {Yusof}, {Even}, {Fryer}, {Stiavelli}, {Chen}, \& {Joggerst}}]{Whalen2014}
{Whalen}, D.~J., {Smidt}, J., {Heger}, A., {et~al.} 2014,
  \href{http://dx.doi.org/10.1088/0004-637X/797/1/9}{\JournalTitle{\apj}, 797,
  9}

\bibitem[{{Woosley} {et~al.}(2007){Woosley}, {Blinnikov}, \&
  {Heger}}]{Woosley2007}
{Woosley}, S.~E., {Blinnikov}, S., \& {Heger}, A. 2007,
  \href{http://dx.doi.org/10.1038/nature06333}{\JournalTitle{\nat}, 450, 390}

\bibitem[{{Yoon} {et~al.}(2012){Yoon}, {Dierks}, \& {Langer}}]{YDL12}
{Yoon}, S.-C., {Dierks}, A., \& {Langer}, N. 2012,
  \href{http://dx.doi.org/10.1051/0004-6361/201117769}{\JournalTitle{\aap},
  542, A113}

\bibitem[{{Yoshida} {et~al.}(2008){Yoshida}, {Omukai}, \&
  {Hernquist}}]{Yoshida2008}
{Yoshida}, N., {Omukai}, K., \& {Hernquist}, L. 2008,
  \href{http://dx.doi.org/10.1126/science.1160259}{\JournalTitle{Science}, 321,
  669}

\bibitem[{{Yoshida} \& {Umeda}(2011)}]{YH11}
{Yoshida}, T., \& {Umeda}, H. 2011,
  \href{http://dx.doi.org/10.1111/j.1745-3933.2011.01008.x}{\JournalTitle{\mnras},
  412, L78}

\bibitem[{{Yusof} {et~al.}(2013){Yusof}, {Hirschi}, {Meynet}, {Crowther},
  {Ekstr{\"o}m}, {Frischknecht}, {Georgy}, {Abu Kassim}, \&
  {Schnurr}}]{Yusof2013}
{Yusof}, N., {Hirschi}, R., {Meynet}, G., {et~al.} 2013,
  \href{http://dx.doi.org/10.1093/mnras/stt794}{\JournalTitle{\mnras}, 433,
  1114}

\end{thebibliography}

\end{document}